\documentclass[11pt]{article}
\usepackage{fullpage,epsf,amssymb,amsthm,amsfonts,amsmath,latexsym,array,graphicx,extarrows,mathtools,mathstyle,mathrsfs,slashed,subcaption,amsbsy,csquotes,accents}
\usepackage[normalem]{ulem}
\usepackage{authblk}
\usepackage{cite}
\usepackage{color}
\usepackage{xcolor}
\usepackage{float} 
\usepackage{hyperref}
\usepackage[export]{adjustbox}
\usepackage{hhline}

\numberwithin{equation}{section}

\title{\bf{Pseudo-scalar meson spectral properties in the chiral crossover region of QCD}} 

\author[1]{Dibyendu Bala}
\author[1]{Olaf Kaczmarek}
\author[2]{Peter Lowdon\thanks{Corresponding author: lowdon@itp.uni-frankfurt.de}}
\author[2,3]{Owe Philipsen}
\author[1]{Tristan Ueding}

\affil[1]{{\scriptsize Fakult\"{a}t f\"{u}r Physik, Universit\"{a}t Bielefeld, 33615 Bielefeld, Germany}}
\affil[2]{{\scriptsize Institut f\"{u}r Theoretische Physik, Goethe-Universit\"{a}t, Max-von-Laue-Str. 1,  60438 Frankfurt am Main, Germany}}
\affil[3]{{\scriptsize John von Neumann Institute for Computing (NIC) at GSI, Planckstr.\ 1, 64291 Darmstadt, Germany}}

\date{}
\begin{document}
\maketitle

\begin{abstract}
\noindent
Determining the type of excitations that can exist in a thermal medium is key to understanding how hadronic matter behaves at extreme temperatures. In this work we study this question for pseudo-scalar mesons comprised of light-strange and strange-strange quarks, analysing how their low-energy spectral properties are modified as one passes through the high-temperature chiral crossover region between $T=145.6  \, \text{MeV}$ and $172.3 \, \text{MeV}$. We utilise the non-perturbative constraints satisfied by correlation functions at finite temperature in order to extract the low-energy meson spectral function contributions from spatial correlator lattice data in $N_{f}=2+1$ flavour QCD. The robustness of these contributions are tested by comparing their predictions with data for the corresponding temporal correlator at different momentum values. We find that around the pseudo-critical temperature $T_{\text{pc}}$ the data in both the light-strange and strange-strange channels is consistent with the presence of a distinct stable particle-like ground state component, a so-called thermoparticle excitation. As the temperature increases this excitation undergoes collisional broadening, and this is qualitatively the same in both channels. These findings suggest that pseudo-scalar mesons in QCD have a bound-state-like structure at low energies within the chiral crossover region which is still strongly influenced by the vacuum states of the theory. 
 
\end{abstract}

\newpage

\section{Introduction}
\label{intro}

The nature of excitations in a thermal medium is an open question that underpins many physical phenomena. In order to address this question it is essential to determine the non-perturbative structure of correlation functions in quantum field theories (QFTs) at finite temperature, since these encode the in-medium dynamics. Significant advancements in the understanding of strongly-interacting phenomena at finite temperatures have been achieved by using lattice quantum chromodynamics (QCD) simulations to compute the form of correlation functions, and to infer dynamical features based on the properties of these quantities~\cite{Detar:1987kae,Born:1991zz,Florkowski:1993bq,Kogut:1998rh,Aarts:2005hg,Wetzorke:2001dk,Karsch:2003jg,Petreczky:2003iz,Asakawa:2003re,Aarts:2006em, Ding:2012sp,Burnier:2015tda,Mukherjee:2015mxc,Meyer:2017ydp,Rothkopf:2019ipj}. Lattice simulations are performed in imaginary time $\tau$, and the correlation functions of particular interest for the study of spectral properties at finite temperature are the two-point functions of gauge-invariant operators $O_{\Gamma}(\tau,\vec{x})$
\begin{align}
C_{\Gamma}(\tau,\vec{x}) = \langle O_{\Gamma}(\tau,\vec{x})\,O_{\Gamma}^{\dagger}(0,\vec{0})\rangle_{\beta},
\label{corrT}
\end{align}
where $\Gamma$ denotes a specific set of quantum numbers, and the expectation value is with respect to the thermal background state at temperature $T=1/\beta$. Assuming that the background state is in thermal equilibrium, it follows from taking the spatial Fourier transform of Eq.~\eqref{corrT} that~\cite{Kapusta:2006pm,Bellac:2011kqa} 
\begin{align}
\widetilde{C}_{\Gamma}(\tau,\vec{p}) = \int_{0}^{\infty} \frac{d\omega}{2\pi} \frac{\cosh\left[\left(\frac{\beta}{2}-|\tau| \right)\omega\right] }{\sinh\left(\frac{\beta}{2}\omega\right)} \,\rho_{\Gamma}(\omega,\vec{p}), 
\label{corrTp}
\end{align}
where $\rho_{\Gamma}(\omega,\vec{p})$ is the Fourier transform of the thermal commutator of $O_{\Gamma}(\tau,\vec{x})$. The spectral function $\rho_{\Gamma}(\omega,\vec{p})$ is a quantity of particular importance as it contains information about all possible spectral excitations with quantum number $\Gamma$ that can occur in the thermal ground state. Equation~\eqref{corrTp} therefore implies that imaginary-time \textit{temporal correlator} $\widetilde{C}_{\Gamma}(\tau,\vec{p})$ data can in principle be used to extract spectral information. This pursuit has received significant focus in the literature but requires one to overcome an ill-posed inverse problem, namely that discrete temporal correlator data and the structure of Eq.~\eqref{corrTp} are not sufficient to uniquely reconstruct the form of $\rho_{\Gamma}(\omega,\vec{p})$. For this reason, all of the strategies for obtaining information about $\rho_{\Gamma}(\omega,\vec{p})$ require additional input, either from perturbative calculations or phenomenological modelling~\cite{Asakawa:2000tr,Meyer:2011gj}. \\

\noindent 
In the specific case $\vec{p}=0$, the temporal correlator $\widetilde{C}_{\Gamma}(\tau,\vec{p})$ coincides with the spatial integral over $C_{\Gamma}(\tau,\vec{x})$. If one instead fixes a spatial direction $z$, and integrates $C_{\Gamma}(\tau,\vec{x})$ over $\tau$ and the remaining directions $x$ and $y$, one obtains the so-called \textit{spatial correlator} $C_{\Gamma}(z)$. For hadronic systems, particular progress has been made in establishing their properties in recent years~\cite{Laermann:2001vg,Wissel:2005pb,Cheng:2010fe,Banerjee:2011yd,Karsch:2012na,Brandt:2014uda,Bazavov:2014cta,Bazavov:2019www,DallaBrida:2021ddx}. At large distances $z$, it is expected that these correlators have an exponential behaviour
\begin{align}
C_{\Gamma}(z) \sim  e^{- \, m_{\Gamma}^{\text{scr}} z},
\end{align}
where the coefficient $m_{\Gamma}^{\text{scr}}$ is referred to as the \textit{screening mass}. In the zero-temperature limit $m_{\Gamma}^{\text{scr}}$ approaches the mass of the lowest-energy vacuum state created by $O_{\Gamma}$, and so in this sense the screening mass provides a measure of the in-medium modifications experienced by this state. A significant advantage of spatial as opposed to temporal correlators is that they can be calculated on the lattice at arbitrarily large distances, whereas the temporal correlator extent is limited by the inverse temperature $\beta$ due to its periodicity. It is therefore computationally easier to extract information about the thermal properties of a system from $C_{\Gamma}(z)$, particularly at large temperatures. As with the temporal correlator, $C_{\Gamma}(z)$ is also directly related to $\rho_{\Gamma}(\omega,\vec{p})$. In this case, one finds
\begin{align}
C_{\Gamma}(z) = \int_{-\infty}^{\infty}  \frac{dp_{z}}{2\pi}e^{i p_{z} z} \int_{0}^{\infty}  \frac{d\omega}{\pi \omega}  \ \rho_{\Gamma}(\omega,p_{x}=p_{y}=0,p_{z}). 
\label{C_rho}
\end{align}
The fact that $C_{\Gamma}(z)$ and $\widetilde{C}_{\Gamma}(\tau,\vec{p})$ are both governed by $\rho_{\Gamma}(\omega,\vec{p})$, but the dependence is significantly different, means that lattice simulations of these correlators provide distinct ways in which to probe the spectral structure of the theory. Therefore, any extraction of $\rho_{\Gamma}(\omega,\vec{p})$ from lattice correlator data must be consistent with \textit{both} sets of data. \\

\noindent
In Ref.~\cite{Lowdon:2022xcl} it was demonstrated that spectral function components can also be extracted from spatial correlator data. This approach was applied to $N_{f}=2$ flavour lattice QCD data for the light-quark pseudo-scalar meson correlator at different temperatures above the pseudo-critical temperature $T_{\!\text{pc}}$, and used to extract the low-energy contributions to $\rho_{\text{PS}}(\omega,\vec{p})$. By computing the temporal correlator predictions from these components via Eq.~\eqref{corrTp}, it was found that the corresponding temporal correlator data does indeed impose non-trivial constraints on these components. The goal of the present work is to generalise the approach developed in Ref.~\cite{Lowdon:2022xcl} to $N_{f}=2+1$ flavour lattice data for pseudo-scalar meson correlators involving both one and two strange-quark components. The remainder of this paper is structured as follows: in Sec.~\ref{spatial_corr} we outline the theoretical foundations of the approach developed in Ref.~\cite{Lowdon:2022xcl}, in Sec.~\ref{PS_analysis} we discuss the lattice data analysis and physical implications of the results, and finally in Sec.~\ref{concl} we summarise our findings.

\section{Spatial correlators at finite temperature}
\label{spatial_corr}

As outlined in Sec.~\ref{intro}, spatial correlators $C(z)$ contain important information about the properties of QFTs at finite temperature. In Ref.~\cite{Lowdon:2022xcl} it was shown that the spatial correlator actually possesses a general non-perturbative representation, the structure of which encodes the connection between the spectral degrees of freedom and the behaviour of $C(z)$. In this section we will briefly outline the origin of this representation, and how it can be used to extract spectral information from spatial correlator data.

\subsection{Causality constraints}
\label{causal_constr}
 
In Refs.~\cite{Bros:1992ey,Buchholz:1993kp,Bros:1995he,Bros:1996mw,Bros:2001zs,Bros:1998ua} the authors developed a non-perturbative framework for describing the structure of scalar correlation functions at finite temperature, generalising the rigorous vacuum-state QFT formulations that have led to numerous foundational insights over the last few decades~\cite{Streater:1989vi,Haag:1992hx,Bogolyubov:1990kw}. In particular, it was demonstrated that the constraint of causality, namely that the fields $\phi(x)$ satisfy the condition: $\left[\phi(x),\phi(y)\right]=0$ for $(x-y)^{2}<0$, imposes significant constraints on the structure of thermal correlation functions. In the case of the spectral function $\rho(\omega,\vec{p})$, this constraint implies that $\rho(\omega,\vec{p})$ satisfies the general representation~\cite{Bros:1992ey}
\begin{align}
\rho(\omega,\vec{p}) = \int_{0}^{\infty} \! ds \int \! \frac{d^{3}\vec{u}}{(2\pi)^{2}} \ \epsilon(\omega) \, \delta\!\left(\omega^{2} - (\vec{p}-\vec{u})^{2} - s \right)\widetilde{D}_{\beta}(\vec{u},s),
\label{commutator_rep}
\end{align}
where the quantum number label $\Gamma$ is dropped here to indicate that this holds for any type of scalar field $\phi(x)$. Equation~\eqref{commutator_rep} corresponds to the finite-temperature generalisation of the K\"{a}ll\'{e}n-Lehmann spectral representation~\cite{Kallen:1952zz,Lehmann:1954xi}, a foundational result of vacuum-state QFT. A significant characteristic of Eq.~\eqref{commutator_rep} is that \textit{all} of the dynamical and temperature-dependent effects are encoded in $\widetilde{D}_{\beta}(\vec{u},s)$, the so-called \textit{thermal spectral density}, and this controls the correlation between the energy and momentum of the system. Establishing the basic properties of $\widetilde{D}_{\beta}(\vec{u},s)$ is therefore key to understanding in-medium phenomena at finite temperature. For the purposes of this analysis we are interested in the general structure of the spatial correlator $C(z)$, as defined in Eq.~\eqref{C_rho}. It turns out that Eq.~\eqref{commutator_rep} imposes various constraints on the structure of the corresponding Euclidean two-point function~\cite{{Lowdon:2022keu}}, and this implies that $C(z)$ can be written~\cite{Lowdon:2022xcl}   
\begin{align}
C(z) = \frac{1}{2}\int_{0}^{\infty} \! ds \int^{\infty}_{|z|} \! dR \ e^{-R\sqrt{s}} D_{\beta}(R,s),
\label{C_int}
\end{align}  
where $D_{\beta}(R,s)=D_{\beta}(|\vec{x}|=R,s)$ is the position-space thermal spectral density, which only depends on $|\vec{x}|$ due to the assumed isotropy of the thermal ground state. Since the properties of the spectral function are entirely determined by the structure of $D_{\beta}(\vec{x},s)$, it is not surprising that this is also the case for $C(z)$. Since Eq.~\eqref{C_int} is a general representation, essentially relying only on the causality of the theory, one can use it to investigate how model-specific characteristics are reflected in the properties of $C(z)$. This will be discussed in Sec.~\ref{spec_prop}.

\subsection{Spectral properties from spatial correlators}
\label{spec_prop}

It is clear that establishing the properties of the thermal spectral density is key for understanding how in-medium effects manifest themselves in correlation functions. In the limit of vanishing temperature full Lorentz invariance is restored, which implies:
\begin{align}
\widetilde{D}_{\beta}(\vec{u},s) \xrightarrow{\beta \rightarrow \infty} (2\pi)^{3} \delta^{3}(\vec{u})\, \rho(s),
\end{align}
or equivalently: $D_{\beta}(\vec{x},s) \rightarrow \rho(s)$, where $\rho(s)$ is the spectral density of the zero-temperature theory. As is well known, the analytic properties of the two-point correlation functions are then fixed by $\rho(s)$, the singularities of which reflect the existence of particle states in the spectrum of the theory. In particular, the appearance of a singular $\delta(s-m^{2})$ component implies the presence of an on-shell state with mass $m$. At vanishing temperature, the spectral characteristics of the theory are directly related to the singular structure of $\rho(\omega,\vec{p})$ in the variable $p^{2}=\omega^{2}-|\vec{p}|^{2}$. However, as soon as the temperature becomes non-vanishing $\widetilde{D}_{\beta}(\vec{u},s)$ will possess a non-trivial $\vec{u}$ dependence, and so the connection between the structure of $\rho(\omega,\vec{p})$ and the spectral properties of the theory is less clear cut. Establishing this connection requires one to understand the dispersive properties of the excitations of the thermal ground state. 

\subsubsection{Thermoparticles}

In Ref.~\cite{Bros:1992ey} it was suggested that the thermal medium may contain particle-like excitations. A characteristic feature of these excitations is that they are formed from the same discrete $\delta(s-m^{2})$ component as mass $m$ vacuum particle states, and their contribution to the position-space thermal spectral density is given by~\cite{Bros:2001zs}
\begin{align}
D_{\beta}(\vec{x},s)= D_{m,\beta}(\vec{x})\, \delta(s-m^{2}) + D_{c, \beta}(\vec{x},s),
\label{decomp}
\end{align}
where the second term $D_{c, \beta}(\vec{x},s)$ parametrises all other spectral contributions. These excitations were subsequently given the name \textit{thermoparticles}~\cite{Buchholz:1993kp} in order to draw a sharp distinction with other particle-like contributions considered in the literature such as collective \textit{quasi-particle} modes, which vanish when $T\rightarrow 0$. As outlined in Refs.~\cite{Bros:1992ey,Buchholz:1993kp,Bros:1995he,Bros:1996mw,Bros:2001zs} and summarised recently~\cite{Lowdon:2022yct}, there are several well-motivated reasons for why thermoparticle components $D_{m,\beta}(\vec{x})\, \delta(s-m^{2})$ provide a natural description for particle-like states at finite temperature. Perhaps the most compelling is that due to the representation in Eq.~\eqref{commutator_rep}, the appearance of a non-trivial \textit{damping factor} $D_{m,\beta}(\vec{x})$ causes the zero-temperature peak in $\rho(\omega,\vec{p})$ at $p^{2}=m^{2}$ to become broadened, since the momentum space factor $\widetilde{D}_{m,\beta}(\vec{u})$ is no longer restricted to the point $\vec{u}=0$. Physically, this broadening describes the effect of collisions with the medium, and has been shown to depend on the underlying dynamics of the theory~\cite{Bros:2001zs}, as one would expect. \\

\noindent
Another significant feature of thermoparticles is that they describe intrinsically \textit{stable} excitations which dominate the large-time evolution of the thermal state~\cite{Bros:2001zs}. The probability for their detection is therefore reduced only by there collisions with the medium. In the zero-temperature limit these collisions become increasingly suppressed, and one recovers the original vacuum-state contribution to the spectral function. The factorised structure of the thermoparticle representation also means that it can be directly generalised to describe intrinsically \textit{unstable} excitations, by replacing $\delta(s-m^{2})$ in Eq.~\eqref{decomp} with a suitable resonance-type function, such as a relativistic Breit-Wigner. This description then enables one to distinguish between reductions in particle-detection probability that arise from two completely different physical origins: 

\begin{enumerate}

\item \textbf{Statistical effects} -- caused by collisions with the medium due to the non-vanishing temperature. These effect are parametrised by the damping factor $D_{m,\beta}(\vec{x})$.

\item \textbf{Mixing effects} -- brought about by decay processes that occur due to the intrinsic instability of underlying vacuum states.

\end{enumerate}

We will not discuss further here the specific structure of thermoparticle contributions and their corresponding damping factors, only to mention that their properties have been explored in different models~\cite{Bros:2001zs}, and that more recently in Refs.~\cite{Lowdon:2021ehf} and~\cite{Lowdon:2022ird} it was shown that these quantities can in fact be used to perform non-perturbative analytic calculations of different in-medium observables, including the shear viscosity.

\subsubsection{Thermoparticle signatures}
\label{themo_sign}

Now that we have a physically well-motivated description of how stable particle-like excitations could potentially appear at finite temperature, one can use this to look for their signatures in thermal correlation functions. In the case of the spectral function, it follows from Eq.~\eqref{commutator_rep} that the corresponding thermoparticle contribution $\rho_{\text{TP}}(\omega,\vec{p})$ always has a discrete energy threshold at the zero-temperature particle mass $m$, and thus one can write
\begin{align}
\rho_{\text{TP}}(\omega,\vec{p}) = \theta(\omega^{2}-m^{2}) \varrho(\omega,\vec{p}),
\label{spec_TP}
\end{align}
where the structure of $\varrho$ depends on the specific properties of the damping factor. From a physical perspective, Eq.~\eqref{spec_TP} implies that the thermal background must be supplied with an energy of at least $m$ in order to have a non-vanishing probability of creating a thermoparticle state. The threshold in Eq.~\eqref{spec_TP} is in fact a special case of a more general situation: 
\begin{align*}
\textit{If $D_{\beta}(\vec{x},s)$ has a threshold at $s=s^{*}$ \ $\Longrightarrow$ \ $\rho(\omega,\vec{p})$ has thresholds when $|\omega|= \sqrt{s^{*}}$.}  
\end{align*}
Of course, given the decomposition in Eq.~\eqref{decomp} the full spectral function would also contain additional components from $D_{c, \beta}(\vec{x},s)$, and this could give spectral contributions both above and below the thermoparticle threshold $|\omega|= m$. The interplay between these different spectral contributions and their relative size depends on the dynamics of the specific theory.   \\

\noindent
For the spatial correlator $C(z)$ it was shown in Ref.~\cite{Lowdon:2022xcl} that if thermoparticle components dictate the low-energy behaviour of the spectral function, just as stable particle states do at zero temperature, then these contributions will dominate the behaviour of the spatial correlator in the following manner
\begin{align}
C(z) \approx \frac{1}{2}  \int^{\infty}_{|z|} \! dR \  e^{-m R} D_{m,\beta}(R),
\label{C_decomp_dom}
\end{align}
and this domination will be especially pronounced at large distances. Taking the derivative of Eq.~\eqref{C_decomp_dom}, one finds that the damping factor can then be extracted from the leading large-$z$ behaviour of the spatial correlator derivative
\begin{align}
D_{m,\beta}(|\vec{x}|=z) \sim -2 \, e^{m z} \, \frac{d C(z)}{dz}, \quad\quad z \rightarrow \infty. 
\label{D_C_rel}
\end{align}
Given the form of the damping factor, one can then use the spectral representation in Eq.~\eqref{commutator_rep} to calculate the thermoparticle contribution to $\rho(\omega,\vec{p})$. This procedure can also be generalised to situations in which there exists more than one thermoparticle-type state~\cite{Lowdon:2022xcl}.    \\

\noindent
The discussions in this section demonstrate that because of the constraints imposed by causality, if thermoparticles exist and provide a dominant low-energy spectral function contribution, their complete analytic structure can be computed \textit{directly} from the corresponding spatial correlator. In Ref.~\cite{Lowdon:2022xcl} this procedure was applied to QCD in order to explore the thermal properties of light meson states. This study focussed in particular on two-flavour lattice QCD data for the spatial pseudo-scalar meson correlator, and found evidence for the existence of a pion ground state consistent with a thermoparticle-type behaviour at a temperature $220\, \text{MeV}$ above the pseudo-critical temperature $T_{\text{pc}}$. A natural question is whether this type of spectral characteristic is \textit{also} displayed in meson states comprised of heavier quarks. In Sec.~\ref{PS_analysis} we will explore and confirm this possibility using lattice QCD data for $N_{f}=2+1$ flavour pseudo-scalar meson correlators.

\section{Pseudo-scalar meson correlator analysis}
\label{PS_analysis}

\subsection{Lattice setup and analysis strategy}
\label{lattice_setup}

For the analysis in this work we studied the Euclidean correlators of pseudo-scalar meson operators $O_{\text{PS}}$ in QCD, focussing in particular on the meson channels comprised of one light and one strange quark: $O_{\text{PS}}= \bar{l}\gamma_{5}s$, and two strange quarks: $O_{\text{PS}}= \bar{s}\gamma_{5}s$. For this purpose we calculated both spatial and temporal correlators in $N_{f}=2+1$ flavour QCD with two mass-degenerate flavours of light quarks and a strange quark, in two different configurations: $\left\{\beta = 7.010, \ T=36.4,  145.6 \, \text{MeV}\right\}$, and $\left\{\beta = 7.188, \ T=43.1,  172.3 \, \text{MeV}\right\}$. We used gauge field configurations generated with the highly improved staggered quark (HISQ) action with physical values for the light and strange quark masses. Details of these configurations, including the choice of scale setting, can be found in Ref.~\cite{Bazavov:2019www}. The logic for choosing these configurations is that it enabled a self-consistent computation of the correlators at an effectively vanishing temperature and at temperatures both below and above the pseudo-critical temperature $T_{\text{pc}}=156.5 \, \text{MeV}$~\cite{HotQCD:2018pds}, hence allowing one to investigate meson spectral properties within the crossover region. The calculation of the correlators were performed using M\"{o}bius domain wall fermions (MDWF)~\cite{Brower:2012vk} and a mixed action approach, with domain wall valence quarks and staggered sea quarks. More details regarding the numerical setup, including the choice of lattice action and parameter tunings, can be found in Appendix~\ref{AppendixMixedActionTuning}. The analysis strategy was based on the proceedure adopted in Ref.~\cite{Lowdon:2022xcl}, and used the following steps:

\begin{enumerate}

\item Fit the form of the spatial correlator $C_{\text{PS}}(z)$ at each temperature. In both channels and in each configuration these data were consistent with a sum of pure exponentials, and so the following functional form was fitted
\begin{align}
C_{\text{PS}}(z) = \sum_{i=1}^{N} c_{i} \left[ e^{-m_{i}^{\text{scr}}z} + e^{-m_{i}^{\text{scr}}(L-z)} \right], 
\label{C_fit}
\end{align}
with $m_{i}^{\text{scr}}$ the screening mass of the different components. The results of these fits are listed in Tables~\ref{tabconf_ls} and~\ref{tabconf_ss}. The second term is needed in order to take into account the periodic boundary conditions of the lattice meson fields along the $z$ direction, with $L$ the symmetric spatial lattice extent. Since $m_{i}^{\text{scr}} \rightarrow m_{i}$ for vanishing temperature, where $m_{i}$ is the corresponding vacuum-state mass, analysis of the low-temperature configuration samples ($36.4 \, \text{MeV}$ and $43.1 \, \text{MeV}$) enabled us to determine an estimate for $m_{i}$.        

\item Use the best-fit parameters $\left\{c_{i},m_{i}^{\text{scr}}\right\}$ at the higher temperatures ($145.6 \, \text{MeV}$ and $172.3 \, \text{MeV}$) and the extracted zero-temperature masses $m_{i}$, together with the analytic relations in Sec.~\ref{spec_prop}, to compute the damping factors of the hypothesised thermoparticle components. As demonstrated in Ref.~\cite{Lowdon:2022xcl}, if these components are present, and dominate at low energies, it follows that the summed pure exponential behaviour in Eq.~\eqref{C_fit} corresponds to $N$ thermoparticle states, each with damping factors of the form
\begin{align}
D_{m_{i},\beta}^{(i)}(\vec{x}) =  \alpha_{i} e^{-\gamma_{i}|\vec{x}|}, \quad\quad \alpha_{i} = 2c_{i}m_{i}^{\text{scr}}, \  \gamma_{i}= m_{i}^{\text{scr}}-m_{i}.
\label{thermo_damping}
\end{align}  
The low-energy contributions from other spectral components in Eq.~\eqref{decomp} are assumed to be negligible in this hypothesis. The validity of this hypothesis is tested in the final step.  

\item Apply the spectral representation in Eq.~\eqref{commutator_rep} to calculate the thermoparticle contributions to the full spectral function $\rho_{\text{PS}}(\omega,\vec{p})$. As with the light-light pseudo-scalar meson case studied in Ref.~\cite{Lowdon:2022xcl}, these contributions $\rho_{\text{PS}}^{(i)}(\omega,\vec{p})$ have the analytic form
\begin{align}
\rho_{\text{PS}}^{(i)}(\omega,\vec{p}) = \epsilon(\omega)  \theta(\omega^{2}-m_{i}^{2}) \,  \frac{4 \, \alpha_{i} \gamma_{i}  \sqrt{\omega^{2}-m_{i}^{2}}}{(|\vec{p}|^{2}+m_{i}^{2}-\omega^{2})^{2} + 2(|\vec{p}|^{2}-m_{i}^{2}+\omega^{2})\gamma_{i}^{2}+\gamma_{i}^{4} }. 
\label{thermo_spec} 
\end{align}

\item Use these spectral function expressions to compute the thermoparticle contribution to the corresponding temporal correlator $\widetilde{C}_{\text{PS}}(\tau,\vec{p})$ via Eq.~\eqref{corrTp}, and compare these predictions with the lattice data. This comparison can be directly made at each temperature by using the non-renormalised lattice data. To compare the spectral functions across different temperatures one must then make a choice of renormalisation prescription.   

\end{enumerate}

\noindent
In step 1 the functional form in Eq.~\eqref{C_fit} was fitted to the lattice data points $C_{\text{PS}}(z=n_{z}a; T)$, with $a$ the lattice spacing and $n_{z}$ the integer spatial site number, where $0 \leq n_{z} \leq N_{s}-1$ with $L= N_{s}a$. These fits were performed in the data range $[n_{\text{min}},N_{s}/2]$ for various values of $n_{\text{min}}$, and for $N=1, 2$ and $3$ states in order to assess their stability. The best-fit parameter values $\left\{c_{i},m_{i}^{\text{scr}}\right\}$ were obtained by minimising the specific Chi-squared statistic $\chi^{2}(n_{\text{min}},N)/\text{d.o.f.}$, and the fits were deemed stable if these values remained insensitive to variations of $n_{\text{min}}$ in some non-trivial range. The final parameter values were determined by averaging the best-fit values over this plateau-like region, and their corresponding errors were obtained using a bootstrap analysis in which fits were performed using random samples of the data in order to estimate the standard deviation. The plotted results of this stability analysis for the best-fit parameters $\left\{c_{i},m_{i}^{\text{scr}}\right\}$ are displayed in Appendix~\ref{stabil_analysis} in Figs.~\ref{ls_145_st}-\ref{ss_172_st}. The horizontal lines in the plots indicate the estimated parameter range from the bootstrap analysis. The plateau-averaged value of the best-fit parameters, and their bootstrap-calculated errors, are listed in Tables~\ref{tabconf_ls} and~\ref{tabconf_ss}. \\

\noindent
Having determined the best estimates for the parameters $\left\{c_{i},m_{i}^{\text{scr}}\right\}$ and their respective errors in step 1, in the remaining steps these values were used to compute the total thermoparticle contributions to $\rho_{\text{PS}}(\omega,\vec{p})$, and compare the corresponding predictions for $\widetilde{C}_{\text{PS}}(\tau,\vec{p})$ with the lattice data. These comparisons are discussed in Secs.~\ref{ls_channel} and~\ref{ss_channel}. Associated with the computations of $\rho_{\text{PS}}(\omega,\vec{p})$ and $\widetilde{C}_{\text{PS}}(\tau,\vec{p})$ are uncertainties due to the error in the fit parameters. These uncertainties were computed using a bootstrap-type analysis and are displayed as error bands in the figures of Secs.~\ref{ls_channel}-\ref{model_discr}. The fit parameter errors also enter in the analysis of the prediction quality in Appendix~\ref{pred_quality}. By following the procedure set out in steps 1--4 we were able to assess whether the QCD lattice data in both the $\bar{l} \gamma_{5} s$ and $\bar{s} \gamma_{5} s$ channels were consistent with the presence of thermoparticle-type components in the spectral function $\rho_{\text{PS}}(\omega,\vec{p})$ both below and above $T_{\text{pc}}$.

\subsection{Light-strange channel}
\label{ls_channel}

Based on the analysis procedure set out in Sec.~\ref{lattice_setup}, the results for the fit parameters and their respective errors are summarised in Table~\ref{tabconf_ls}. For the pseudo-scalar meson operator $O_{\text{PS}}= \bar{l}\gamma_{5}s$ the lowest-lying particle state at zero temperature is the kaon $K$, which is stable in QCD and has a neutral charge state $K^{0}$ of mass $497.6 \, \text{MeV}$~\cite{Workman:2022ynf}. At higher energies the situation is significantly less certain, but a recent analysis from the LHCb experiment suggests that the $K(1460)$ is the next heaviest state, with a mass of $1482.4 \, \text{MeV}$, and corresponds to the first radial excitation of the $K$~\cite{LHCb:2017swu}. From Table~\ref{tabconf_ls} one can see that the low temperature fit for $\beta=7.188$ agrees quite well with the experimentally measured $m_{K}$ and $m_{K(1460)}$ values, but the $\beta=7.010$ case gives an overestimation of the masses, particularly for the excited state. In both configurations there is a significant increase in the screening mass of the kaon between the low and high temperatures, and the values at $145.6 \, \text{MeV}$ and $172.3 \, \text{MeV}$ are consistent with the continuum extrapolated results obtained in Ref.~\cite{Bazavov:2019www}. For the first excited state the change in $m^{\text{scr}}$ runs in opposite directions as the temperature is increased in the different configurations. For $\beta=7.188$ the $m^{\text{scr}}$ value increases significantly, and for $\beta=7.010$ it decreases, although due to the large uncertainties the change between $36.4 \, \text{MeV}$ and $145.6 \, \text{MeV}$ is consistent with zero. \\ 

\begin{table}[t!]
\center
\small
\begin{tabular}{|c|c|c|c|c|c|c|c|c|} 
\hline
\rule{0pt}{3ex}
\!\!\!\! $T \, [\text{MeV}]$ \!\!\!\! & $a \, [\text{fm}]$  &  $\beta$  & $N_{s}^{3} \times N_{\tau}$ & \!\!\! $\text{Conf.} \, \#$ \!\!\! & $c_{1} \, [\text{GeV}^{3}]$ & $m_{1}^{\text{src}} \, [\text{MeV}]$ & $c_{2} \, [\text{GeV}^{3}]$ & $m_{2}^{\text{src}} \, [\text{MeV}]$   \\[0.5ex]
\hhline{|=|=|=|=|=|=|=|=|=|}
\!\!\!\! 36.4  \!\!\!\!  & 0.085 & 7.010  & $64^{3} \times 64$ & \!\!\! 227 \!\!\! & $0.165(5)$ & $502(2)$ & $0.824(362)$ & $2100(215)$ \\ 
 \hline
\!\!\!\! 145.6 \!\!\!\!  & 0.085 & 7.010  & $64^{3} \times 16$ & \!\!\! 399 \!\!\! & $0.171(7)$  & $527(3)$  & $0.497(186)$  & $1732(184)$ \\ 
 \hline
\!\!\!\! 43.1  \!\!\!\!  & 0.072 & 7.188  & $64^{3} \times 64$ & \!\!\! 232 \!\!\! & $0.213(10)$  & $496(2)$  & $0.638(163)$  & $1844(172)$ \\ 
 \hline
\!\!\!\! 172.3 \!\!\!\!  & 0.072 & 7.188  & $64^{3} \times 16$ & \!\!\! 395 \!\!\! & $0.214(11)$  & $632(7)$   & $1.567(571)$  & $2539(266)$ \\ 
 \hline
\end{tabular}
\caption{Parameter fit values and their estimated uncertainties from the $O_{\text{PS}}= \bar{l}\gamma_{5}s$ spatial correlator data.}
\label{tabconf_ls}
\end{table}

\begin{figure}[t!]
\centering
\includegraphics[width=0.49\textwidth]{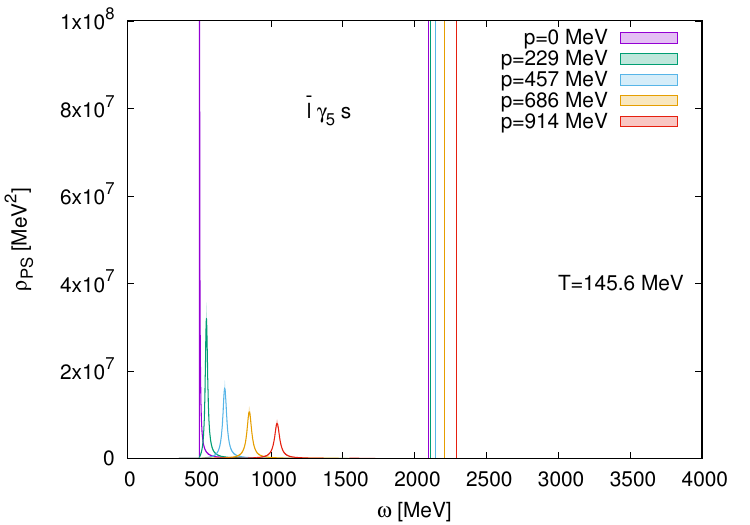}
\includegraphics[width=0.49\textwidth]{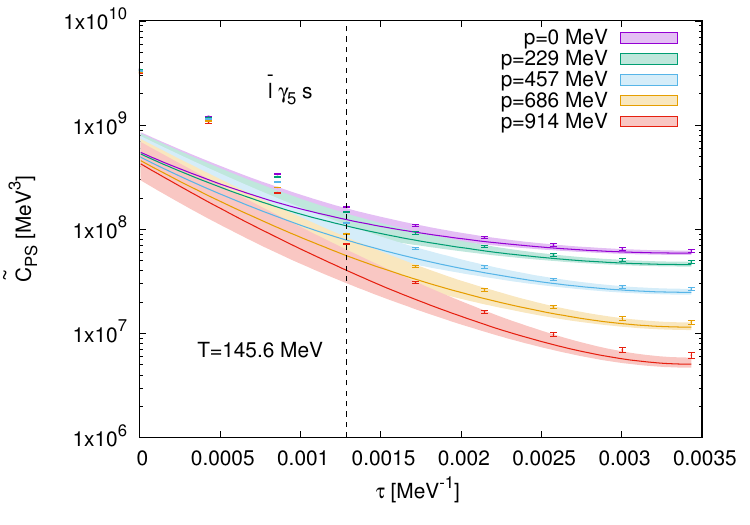}

\caption{$\bar{l} \gamma_{5} s$ channel spectral function extracted from spatial correlator lattice data at $145.6 \, \text{MeV}$ (left), and the temporal correlator prediction and corresponding lattice data (right). The vertical dashed line in the right plot indicates the approximate boundary below which knowledge of further excited states is necessary. The effects of the vacuum mass $m_{i}$ uncertainties are not included in the left plot in order to improve its clarity.}
\label{ls_145_specT}
\end{figure}

\noindent
In Fig.~\ref{ls_145_specT} is displayed the low-energy contribution to the spectral function using the $T=145.6 \, \text{MeV}$ parameters in Table~\ref{tabconf_ls}, together with the resulting prediction for the temporal correlator $\widetilde{C}_{\text{PS}}(\tau,\vec{p})$ and its comparison with the corresponding lattice data for different values\footnote{In this analysis we always assume the lattice data to be approximately isotropic, and take $|\vec{p}|=|p_{x}|$.} of $|\vec{p}|$. In the left plot one can see that the kaon state undergoes collisional broadening but the first excited state remains on its vacuum mass shell. In the right plot one finds that the temporal correlator prediction agrees well\footnote{The extent of this agreement is quantified in Appendix~\ref{pred_quality}.} with the data at large $\tau$, and then starts to underestimate the data as $\tau$ becomes smaller. The point at which this underestimation occurs coincides with the approximate boundary where the two-state spatial correlator fit starts to break down, which is indicated by the vertical dashed line in the right plot. This is consistent with the fact that higher excited states, which should increasingly dominate at smaller $\tau$, are not included in the prediction. It should be noted that for the largest values of $|\vec{p}|$ there start to arise sizeable lattice cutoff effects, and so the comparison at these values becomes significantly less certain. \\

\begin{figure}[t!]
\centering
\includegraphics[width=0.49\textwidth]{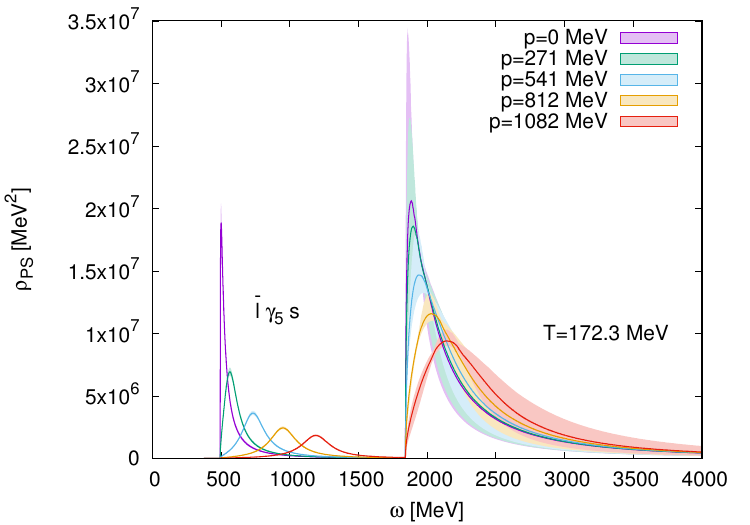}
\includegraphics[width=0.49\textwidth]{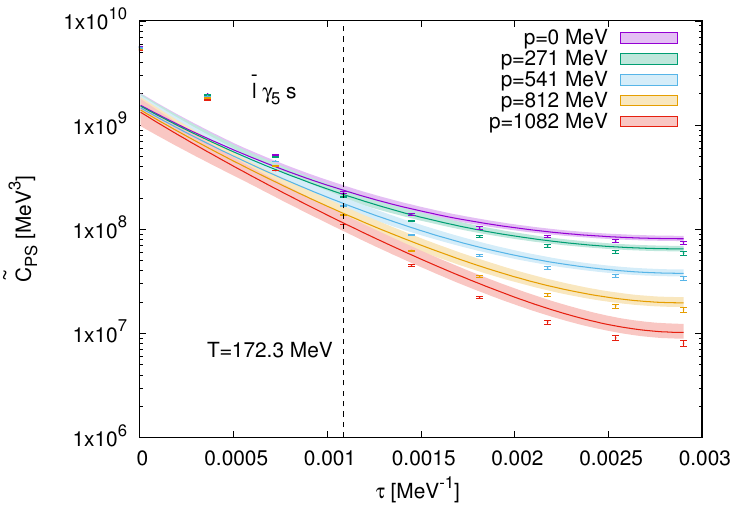}
\caption{$\bar{l} \gamma_{5} s$ channel spectral function extracted from spatial correlator lattice data at $172.3 \, \text{MeV}$ (left), and the temporal correlator prediction and corresponding lattice data (right). The vertical dashed line in the right plot indicates the approximate boundary below which knowledge of further excited states is necessary. The effects of the vacuum mass $m_{i}$ uncertainties are not included in the left plot in order to improve its clarity.}
\label{ls_172_specT}
\end{figure}

\begin{figure}[h!]
\centering
\includegraphics[width=0.49\textwidth]{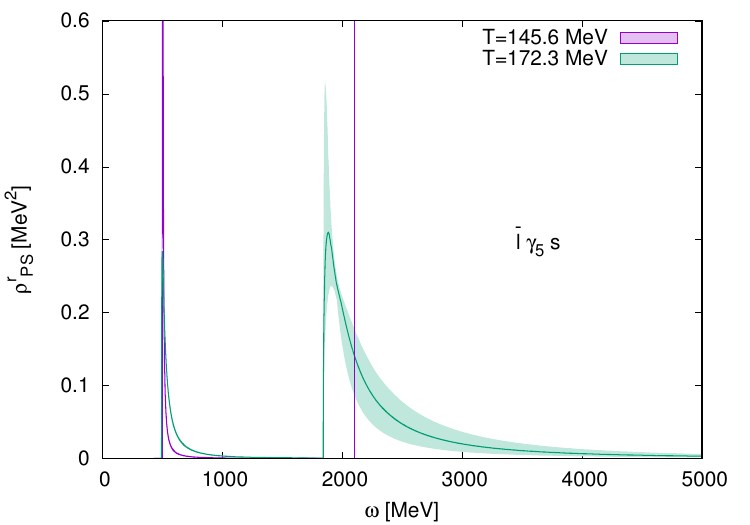}
\caption{Renormalised $\bar{l} \gamma_{5} s$ channel spectral function for $T= 145.6 \, \text{MeV}$ and $172.3 \, \text{MeV}$. The effects of the uncertainties in the vacuum masses $m_{i}$ on the spectral function are not included in the plot. This explains why the excited state components at $T= 145.6 \, \text{MeV}$ and $172.3 \, \text{MeV}$ appear not to possess the same energy threshold, although the extracted vacuum masses are consistent within uncertainties.}
\label{ls_spec_ren}
\end{figure}

\noindent
Figure~\ref{ls_172_specT} shows how the spectral function and corresponding temporal correlator prediction change as the temperature is increased above $T_{\text{pc}}$. One can see that both the kaon and first excited state are significantly broadened, and that the temporal correlator predictions agree with the data, but only for the largest values of $\tau$ and smaller values of momentum $|\vec{p}|$, as quantified in Appendix~\ref{pred_quality}. It is difficult to make firm conclusions here since the deviations at larger $|\vec{p}|$ may be as a result of cutoff effects, and the slight overestimation in prediction between the smallest $\tau$ points and the dashed line, where the two-state fit starts to break down, could be due to systematic effects. As $\tau$ is decreased below the dashed line the prediction immediately begins to undershoot the data, which is consistent with the absence of higher-energy spectral function contributions in the prediction. So far we have considered the form of the non-renormalised spectral functions and corresponding temporal correlator predictions separately for each lattice setup. In order to compare the spectral functions at the temperatures below and above $T_{\text{pc}}$ one must choose a renormalisation prescription. By making the choice: $\widetilde{C}_{\text{PS}}(\tau=0.5 \,\text{fm},\vec{p}=0) = 1 \, \text{MeV}^{3}$ at both $T=36.4 \, \text{MeV}$ and $T=43.1 \, \text{MeV}$, the renormalised spectral function at $\vec{p}=0$ has the form in Fig.~\ref{ls_spec_ren}. One can see that as the temperature increases through the crossover region both the ground and first excited states become collisionally broadened, with the amplitude of their peaks significantly reduced.  \\

\subsection{Strange-strange channel}
\label{ss_channel}

Table~\ref{tabconf_ss} summarises the results obtained for the fit parameters and their respective errors after applying the procedure set out in Sec.~\ref{lattice_setup} to the strange-strange correlator data. Unlike the $\bar{l} \gamma_{5} s$ channel, for the operator $O_{\text{PS}}= \bar{s}\gamma_{5}s$ the lattice correlator can contain a disconnected component, and this could in principle be non-negligible. However, for the purposes of this study we follow the approach of Refs.~\cite{Bazavov:2014cta,Bazavov:2019www} in treating the ground state to be a hypothetical unmixed projection $\eta_{s\bar{s}}$ of the $\eta$ meson which is generated by the purely connected correlator component. The physical characteristics of higher excited states in this channel are less certain, but it has been suggested that $\eta(1295)$ could be the first radial excitation of the $\eta$, and that it is mainly comprised of an $s\bar{s}$ component~\cite{Workman:2022ynf}. In both lattice configurations there is a significant increase in the screening mass of the $\eta_{s\bar{s}}$ between the lower and higher temperature data fits, and the values at all temperatures are consistent with the continuum-extrapolated results obtained in Ref.~\cite{Bazavov:2019www}. For $\beta=7.010$ the screening mass of the first excited state does not change significantly as the temperature is increased, but for $\beta=7.188$ the value of $m^{\text{scr}}$ decreases between $43.1 \, \text{MeV}$ and $172.3 \, \text{MeV}$. \\

\begin{table}[t!]
\center
\small
\begin{tabular}{|c|c|c|c|c|c|c|c|c|} 
\hline
\rule{0pt}{3ex}
\!\!\!\! $T \, [\text{MeV}]$ \!\!\!\! & $a \, [\text{fm}]$  &  $\beta$  & $N_{s}^{3} \times N_{\tau}$ & \!\!\! $\text{Conf.} \, \#$ \!\!\! & $c_{1} \, [\text{GeV}^{3}]$ & $m_{1}^{\text{src}} \, [\text{MeV}]$ & $c_{2} \, [\text{GeV}^{3}]$ & $m_{2}^{\text{src}} \, [\text{MeV}]$   \\[0.5ex]
\hhline{|=|=|=|=|=|=|=|=|=|}
\!\!\!\! 36.4  \!\!\!\!  & 0.085 & 7.010  & $64^{3} \times 64$ & \!\!\! 227 \!\!\! & $0.159(4)$   & $692(1)$   & $0.338(118)$    & $1719(145)$ \\ 
 \hline
\!\!\!\! 145.6 \!\!\!\!  & 0.085 & 7.010  & $64^{3} \times 16$ & \!\!\! 399 \!\!\! & $0.171(4)$   & $714(2)$   & $0.311(100)$    & $1634(116)$ \\ 
 \hline
\!\!\!\! 43.1  \!\!\!\!  & 0.072 & 7.188  & $64^{3} \times 64$ & \!\!\! 232 \!\!\! & $0.200(5)$   & $692(2)$   & $0.617(145)$    & $1967(123)$ \\ 
 \hline
\!\!\!\! 172.3 \!\!\!\!  & 0.072 & 7.188  & $64^{3} \times 16$ & \!\!\! 395 \!\!\! & $0.208(7)$   & $778(3)$   & $0.170(67)$     & $1309(157)$ \\ 
 \hline
\end{tabular}
\caption{Parameter fit values and their estimated uncertainties from the $O_{\text{PS}}= \bar{s}\gamma_{5}s$ spatial correlator data.}
\label{tabconf_ss}
\end{table}

\begin{figure}[t!]
\centering
\includegraphics[width=0.49\textwidth]{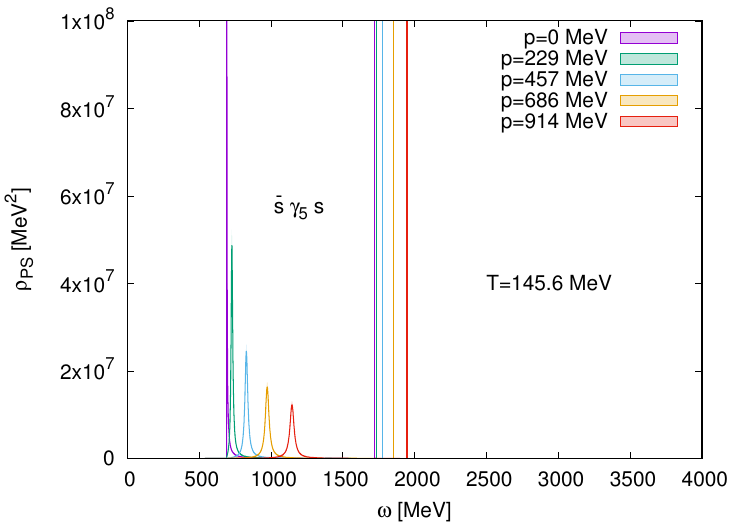}
\includegraphics[width=0.49\textwidth]{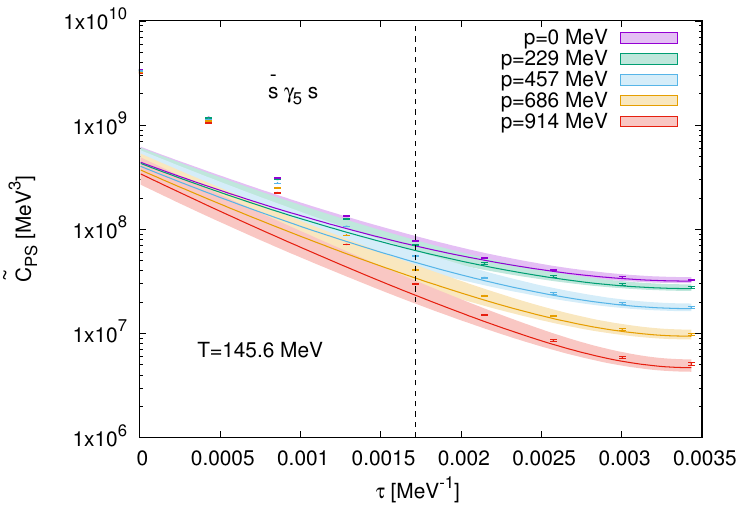}
\caption{$\bar{s} \gamma_{5} s$ channel spectral function extracted from spatial correlator lattice data at $145.6 \, \text{MeV}$ (left), and the temporal correlator prediction and corresponding lattice data (right). The vertical dashed line in the right plot indicates the approximate boundary below which knowledge of further excited states is necessary. The effects of the vacuum mass $m_{i}$ uncertainties are not included in the left plot in order to improve its clarity.}
\label{ss_145_specT}
\end{figure}

\noindent
In Fig.~\ref{ss_145_specT} is displayed the low-energy contribution to the spectral function using the $T=145.6 \, \text{MeV}$ parameters in Table~\ref{tabconf_ss}, together with the resulting prediction for the temporal correlator and its comparison with the corresponding lattice data for different values of $|\vec{p}|$. Just like in the $\bar{l} \gamma_{5} s$ channel, in the left plot one can see that the $\eta_{s\bar{s}}$ ground state broadens, but the first excited state remains on its vacuum mass shell. In the right plot one finds that the temporal correlator prediction also has the same characteristic features, matching the data well\footnote{The extent of this agreement is quantified in Appendix~\ref{pred_quality}.} at large $\tau$, and then starting to increasingly underestimate it as $\tau$ becomes smaller. This coincides with the region where the two-state fits start to break down, and is therefore consistent with the absence of higher-energy spectral function contributions. Although there are sizeable lattice cutoff effects for larger values of momentum, the prediction still provides a good description of the large $\tau$ data for all values of $|\vec{p}|$.  \\

\begin{figure}[t!]
\centering
\includegraphics[width=0.49\textwidth]{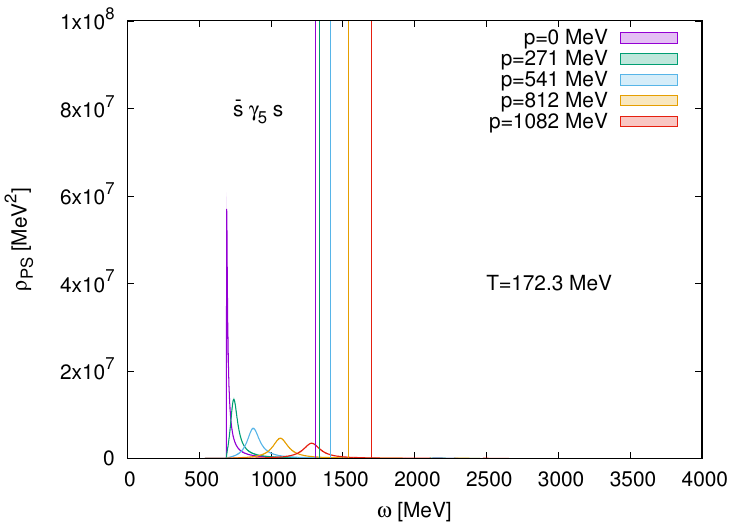}
\includegraphics[width=0.49\textwidth]{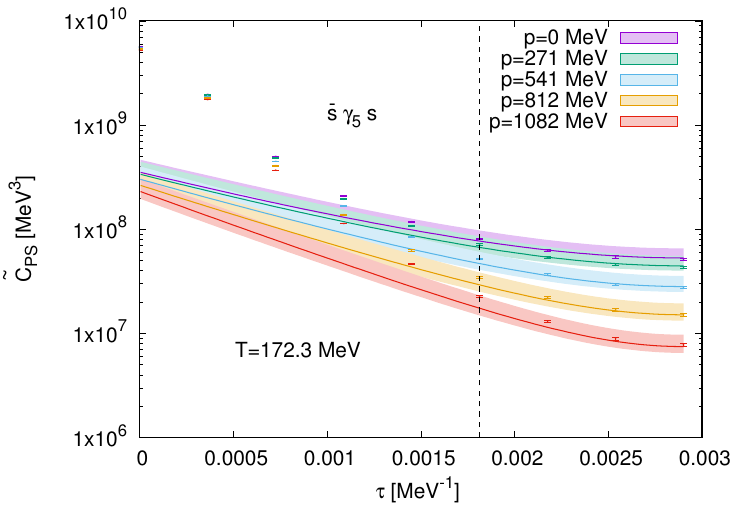}
\caption{$\bar{s} \gamma_{5} s$ channel spectral function extracted from spatial correlator lattice data at $172.3 \, \text{MeV}$ (left), and the temporal correlator prediction and corresponding lattice data (right). The vertical dashed line in the right plot indicates the approximate boundary below which knowledge of further excited states is necessary. In this plot the excited state vacuum mass is set equal to the screening mass. The effects of the vacuum mass $m_{i}$ uncertainties are not included in the left plot in order to improve its clarity.}
\label{ss_172_specT2}
\end{figure}

\begin{figure}[t!]  
\centering
\includegraphics[width=0.49\textwidth]{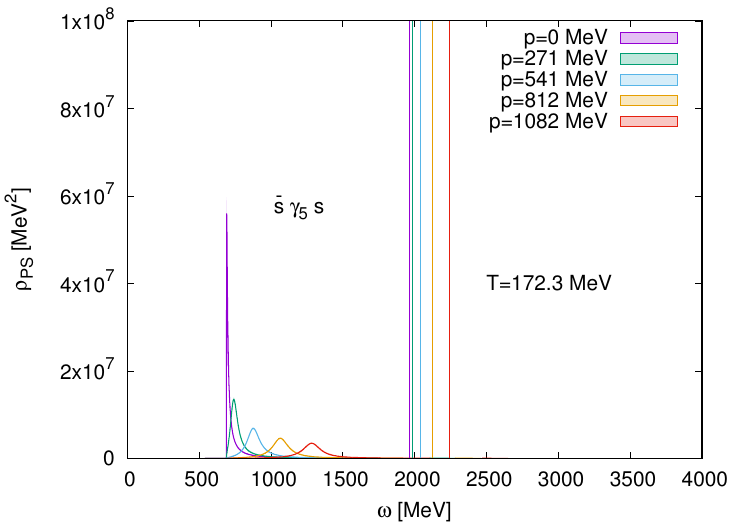}
\includegraphics[width=0.49\textwidth]{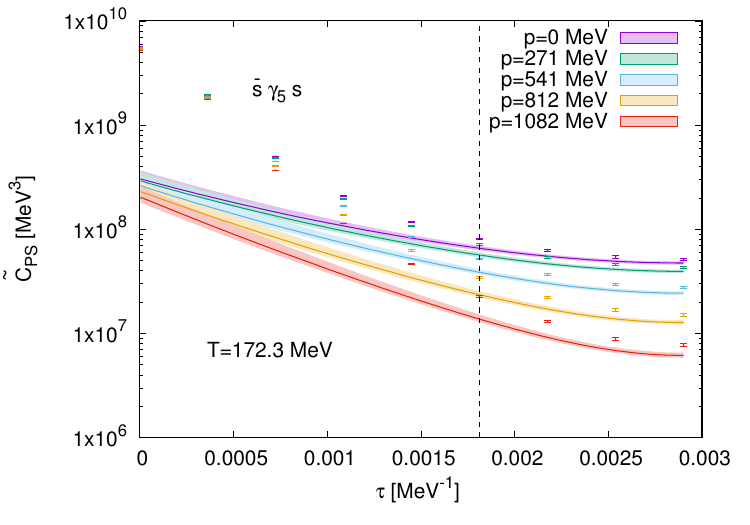}
\caption{$\bar{s} \gamma_{5} s$ channel spectral function extracted from spatial correlator lattice data at $172.3 \, \text{MeV}$ (left), and the temporal correlator prediction and corresponding lattice data (right). The vertical dashed line in the right plot indicates the approximate boundary below which knowledge of further excited states is necessary. In this plot the excited state mass is set equal to its value at $43.1 \, \text{MeV}$. The effects of the vacuum mass $m_{i}$ uncertainties are not included in the left plot in order to improve its clarity.}
\label{ss_172_specT}
\end{figure}

\noindent
Due to the decrease in the excited state screening mass between $43.1 \, \text{MeV}$ and $172.3 \, \text{MeV}$ there are two possible interpretations: either this decrease is not significant once systematic effects are taken into account, or this decrease has a physical origin. Given the first interpretation, one should therefore simply set the vacuum mass of the excited state equal to its screening mass at $172.3 \, \text{MeV}$. The corresponding spectral function and temporal correlator predictions are plotted in Fig.~\ref{ss_172_specT2}. Here one can see very good agreement between the prediction and lattice data at large $\tau$, as is the case at $145.6 \, \text{MeV}$. In the second interpretation, the decrease in the excited state screening mass could be explained by the breakdown in the assumption that the thermoparticle-like components are entirely dominant at low energies. If one fixes the excited state mass to the value obtained at $43.1 \, \text{MeV}$, and assumes that this state does not broaden, one obtains the results in Fig.~\ref{ss_172_specT}. In this case one can see that the predictions underestimate the data at all momentum values. If true, this would give support to the hypothesis that the spectral function contains a significant non-thermoparticle-like low-energy contribution. By definition, this contribution would appear in the thermal spectral density like the second term in Eq.~\eqref{decomp} and ultimately behave like a collective thermal excitation.      
     
\subsection{Model discrimination}
\label{model_discr}

A well-known problem with standard reconstruction approaches is that different reconstructed spectral functions may display large differences, and yet still provide a good description of the correlator data from which they were extracted. However, by comparing both spatial \textit{and} temporal correlator predictions one can break this degeneracy, since the correlators have a different dependence on the spectral function, as discussed in Sec.~\ref{intro}. To demonstrate the utility of this approach, in the remainder of this section we will consider a different causal spectral function which \textit{also} gives rise to a purely exponential spatial correlator contribution, namely the relativistic Breit-Wigner
\begin{align}
\rho_{\text{BW}}(\omega,\vec{p}) = \frac{4  \omega \Gamma}{(\omega^{2}-|\vec{p}|^{2}-m^{2}-\Gamma^{2})^{2} + 4\omega^{2}\Gamma^{2}}, 
\label{rho_BW}
\end{align}    
where $m$ is the zero-temperature mass, and $\Gamma$ a temperature-dependent width. Although this contribution is causal~\cite{Henning:1995ft} it does not describe a thermoparticle-type state, since its corresponding thermal spectral density $D_{\beta}(\vec{x},s)$ cannot be written: $D_{m,\beta}(\vec{x})\, \delta(s-m^{2})$. This implies that a Breit-Wigner excitation of the form in Eq.~\eqref{rho_BW} describes an intrinsically \textit{unstable} state, in contrast to the thermoparticle case. Combining Eqs.~\eqref{rho_BW} and~\eqref{C_rho}, the spatial correlator takes the form
\begin{align}
C_{\text{BW}}(z) = \frac{e^{-\sqrt{m^{2}+\Gamma^{2}}|z|}}{2\sqrt{m^{2}+\Gamma^{2}}}.
\label{C_BW}
\end{align}
One can immediately see that the screening mass of the Breit-Wigner is: $m_{\text{BW}}^{\text{scr}}= \sqrt{m^{2}+\Gamma^{2}}$, which is qualitatively different to the thermoparticle case: $m_{\text{TP}}^{\text{scr}}= m + \gamma$ from Eq.~\eqref{thermo_damping}. \\

\begin{figure}[t!]
\centering
\includegraphics[width=0.49\textwidth]{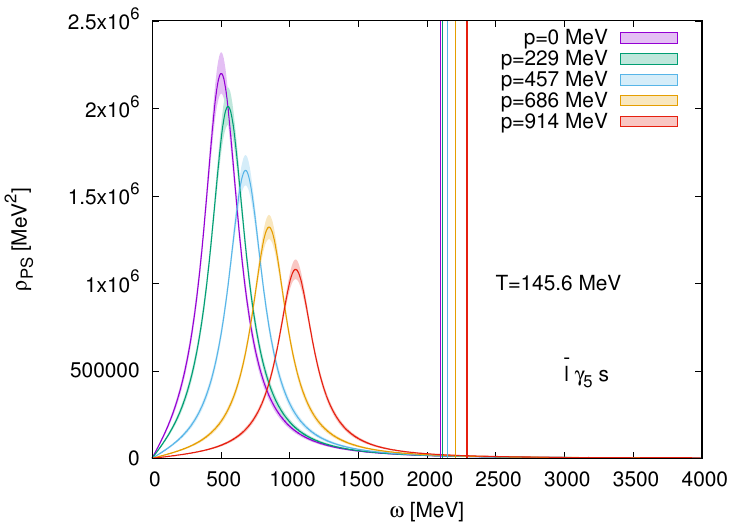}
\includegraphics[width=0.49\textwidth]{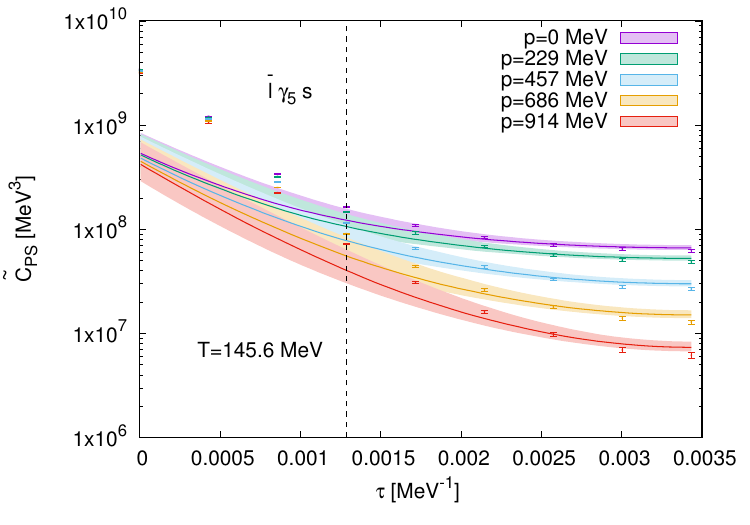}
\caption{$\bar{l} \gamma_{5} s$ spectral function extracted from spatial correlator data at $145.6 \, \text{MeV}$ assuming a Breit-Wigner ground state form (left), and the temporal correlator prediction and corresponding lattice data (right). The vertical dashed line in the right plot indicates the approximate boundary below which knowledge of further excited states is necessary.}
\label{ls_145_BW}
\end{figure}

\begin{figure}[t!]
\centering
\includegraphics[width=0.49\textwidth]{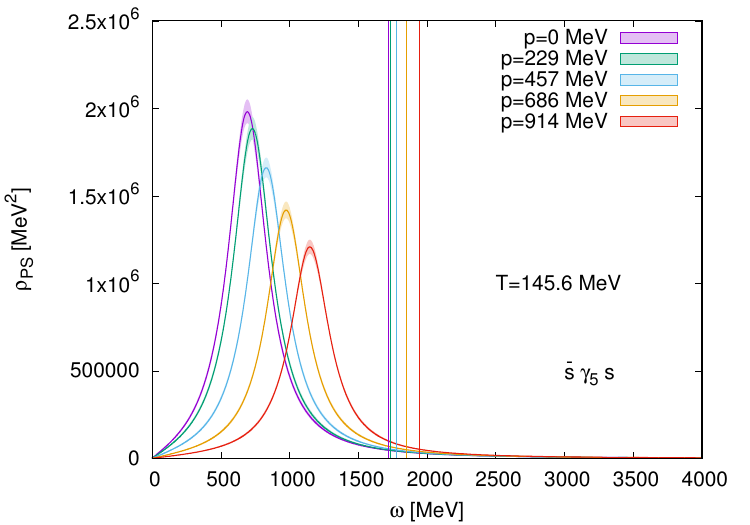}
\includegraphics[width=0.49\textwidth]{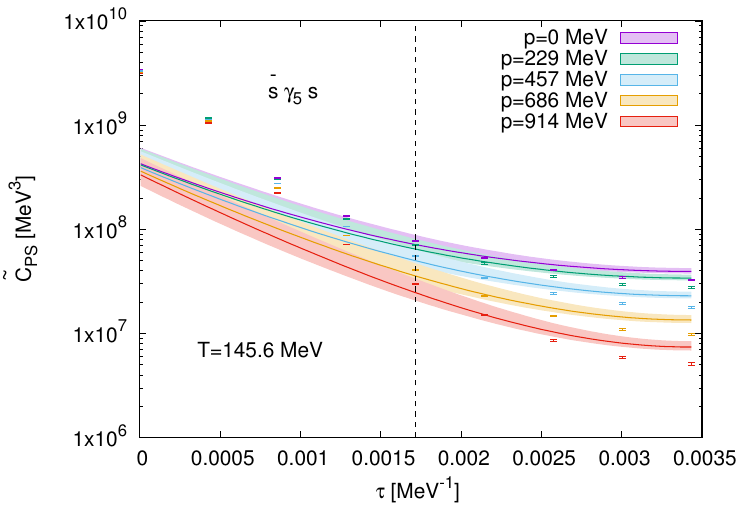}
\caption{$\bar{s} \gamma_{5} s$ spectral function extracted from spatial correlator data at $145.6 \, \text{MeV}$ assuming a Breit-Wigner ground state form (left), and the temporal correlator prediction and corresponding lattice data (right). The vertical dashed line in the right plot indicates the approximate boundary below which knowledge of further excited states is necessary.}
\label{ss_145_BW}
\end{figure}

\begin{figure}[t!]
\centering
\includegraphics[width=0.49\textwidth]{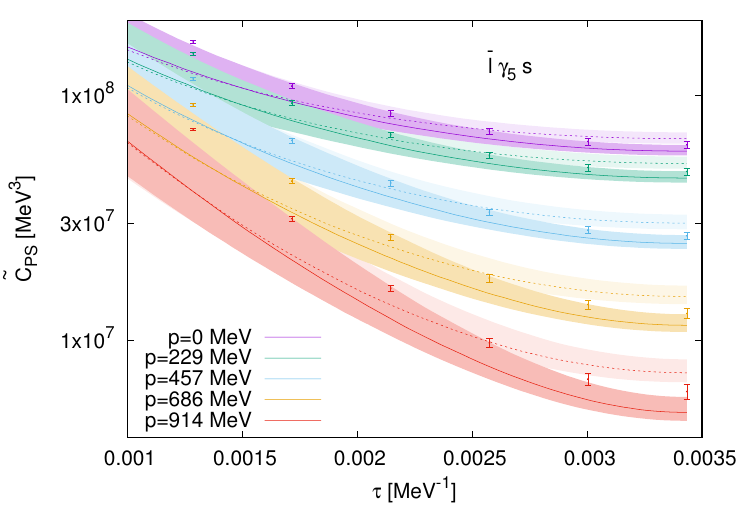}
\includegraphics[width=0.49\textwidth]{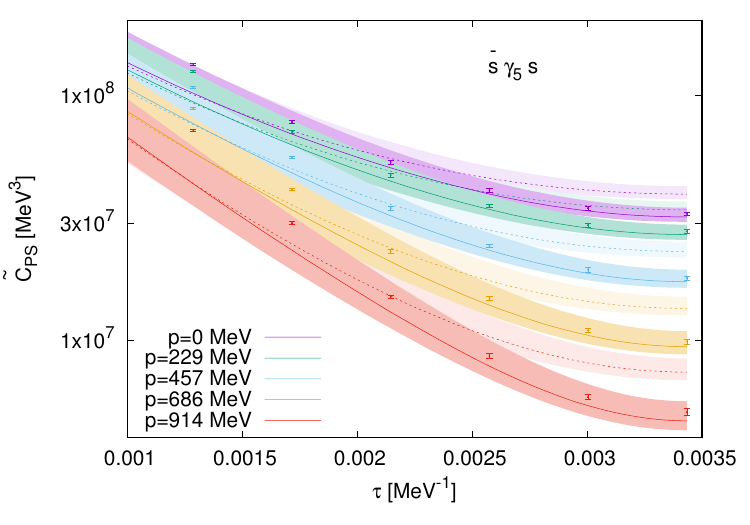}
\caption{Comparison of the temporal correlator predictions with the $145.6 \, \text{MeV}$ lattice data assuming either a thermoparticle (dark shading, solid line) or Breit-Wigner (light shading, dashed line) ground state in the $\bar{l} \gamma_{5} s$ channel (left), and the $\bar{s} \gamma_{5} s$ channel (right). This is plotted in the reduced $\tau$ range for which the ground and first excited states are dominant.}
\label{BW_comp}
\end{figure}

\noindent
With Eq.~\eqref{C_BW} one can now extract the width parameter $\Gamma$ via the screening mass fitted from the lattice data, and use Eqs.~\eqref{corrTp} and~\eqref{C_BW} to predict the form of the corresponding temporal correlator. This is shown in Figs.~\ref{ls_145_BW} and~\ref{ss_145_BW} at the lowest non-trivial temperature $T=145.6 \, \text{MeV}$ in the $\bar{l} \gamma_{5} s$ and $\bar{s} \gamma_{5} s$ channels, respectively. Since the first excited state is consistent with undergoing no broadening at this temperature, these predictions correspond to spectral functions with the same on-shell excited state contribution as the thermoparticle case, but with a Breit-Wigner ground state component. In Figs.~\ref{ls_145_BW} and~\ref{ss_145_BW} there are clear differences with the thermoparticle predictions, which for ease of comparison are plotted together in Fig.~\ref{BW_comp} for the largest $\tau$ data points. The Breit-Wigner contribution gives rise to a flatter temporal correlator behaviour at large $\tau$, and this leads to a poorer description of the data in both meson channels. This is most pronounced in the $\bar{s} \gamma_{5} s$ channel, where the prediction is significantly above the largest $\tau$ data points for all values of $|\vec{p}|$. \\

\noindent
Overall, these findings suggest that the lowest-lying spectral function components in each channel are not consistent with a ground state contribution with a causal Breit-Wigner structure. It is interesting to see here that even at this relatively low temperature of $145.6 \, \text{MeV}$, the comparison of both spatial and temporal correlators results in a non-trivial discriminating power between different spectral function models.

\subsection{Spectral implications for QCD}
\label{QCD_implic}

The analyses in Secs.~\ref{ls_channel} and~\ref{ss_channel} are consistent with the hypothesis that both the $\bar{l} \gamma_{5} s$ and $\bar{s} \gamma_{5} s$ meson spectral functions contain a stable bound-state-like thermoparticle component which dominates the low-energy behaviour of $\rho_{\text{PS}}(\omega,\vec{p})$ at $T=145.6 \, \text{MeV}$, and has the form of Eq.~\eqref{thermo_spec}. In Appendix~\ref{pert_pred} it is demonstrated that these components have a purely non-perturbative character. Although $145.6 \, \text{MeV}$ is a relatively low temperature, the contribution of these components to the temporal correlator is distinct enough to be differentiated from other casual spectral function models, as outlined in Sec.~\ref{model_discr}. There are two particularly significant aspects to these findings:

\begin{enumerate}

\item The thermoparticle picture appears to provide a good description of how particle states are modified in the presence of a thermal medium. Instead of behaving like collective thermal excitations, these states are intrinsically stable, and their amplitudes dissipate solely due to their interactions with the medium.        

\item The qualitative features of $\rho_{\text{PS}}(\omega,\vec{p})$ are the same in \textit{both} meson channels, and in fact also agree with those found in Ref.~\cite{Lowdon:2022xcl} in the $\bar{l} \gamma_{5} l$ channel. This suggests that the low-energy spectral functions of pseudo-scalar mesons in QCD have a common structure which is fixed by the vacuum dynamics of the theory.  

\end{enumerate}

Both of these results give support to the non-perturbative finite-temperature QFT framework put forward in Refs.~\cite{Bros:1992ey,Buchholz:1993kp,Bros:1995he,Bros:1996mw,Bros:2001zs}. The second observation also appears to align with the findings of Ref.~\cite{Bros:2001zs}, where it was demonstrated that the behaviour of damping factors associated with asymptotic states, and hence low-energy spectral components, can actually be fixed by the equations of motion. Due to the significantly larger errors associated with extracting the parameters of excited states, the conclusions regarding the status of these spectral contributions remain more uncertain. But within the estimated uncertainties, the $T=145.6 \, \text{MeV}$ data in both $\bar{l} \gamma_{5} s$ and $\bar{s} \gamma_{5} s$ channels are consistent with the first excited states undergoing no thermal modifications, and hence remaining on their vacuum mass shells. Physically, this is perhaps not so surprising, since the vacuum masses of these states are still significantly larger than the temperature scale, which is below $T_{\text{pc}}$. At $T=172.3 \, \text{MeV}$, which is above $T_{\text{pc}}$, the excited states have more of an effect on the structure of the spectral function. In the $\bar{l} \gamma_{5} s$ channel the results are consistent with the first excited state undergoing collisional broadening, whereas for $\bar{s} \gamma_{5} s$ it appears that the excited state remains on shell, but that there may be additional non-thermoparticle-like contributions. This suggests that as the temperature passes through the crossover region and starts to increase above $T_{\text{pc}}$, the flavour dependence of the meson states has a more significant bearing on the low-energy spectral properties of the theory.

\section{Conclusions} 
\label{concl}

By utilising the general non-perturbative constraints satisfied by correlation functions at finite temperature, in this work we analysed how the spectral properties of pseudo-scalar mesons comprised of light-strange and strange-strange quarks are modified as one passes through the high-temperature chiral crossover region. A key component of this analysis was to test the hypothesis that stable particle-like excitations, so-called \textit{thermoparticles}, exist in thermal media and provide a natural description for low-energy meson states at finite temperature. For this purpose we followed the strategy developed in Ref.~\cite{Lowdon:2022xcl}, where it was established that thermoparticle contributions to spectral functions can be directly extracted from the behaviour of the spatial correlators, and then used to predict the form of the corresponding temporal correlator $\widetilde{C}_{\text{PS}}(\tau,\vec{p})$. Since the spatial and temporal correlators depend on the spectral function in different ways, this provides a highly non-trivial test of the validity of the extracted thermoparticle contributions. \\

\noindent
Applying this procedure to $N_{f}=2+1$ flavour lattice QCD data we show that these data are consistent with the appearance of a stable bound-state-like thermoparticle contribution in the spectral function for both $\bar{l} \gamma_{5} s$ and $\bar{s} \gamma_{5} s$ flavour mesons, and that this contribution is present for temperatures around the pseudo-critical temperature $T_{\text{pc}}$. This generalises the results of Ref.~\cite{Lowdon:2022xcl}, where evidence of such a component was found in the $\bar{l} \gamma_{5} l$ channel for $N_{f}=2$. We tested the robustness of the thermoparticle contribution by comparing the $\widetilde{C}_{\text{PS}}(\tau,\vec{p})$ prediction with lattice data at multiple momentum values, and also with a different causal spectral function model. Overall, these findings suggest that the thermoparticle hypothesis provides a consistent description of how vacuum states undergo collisional broadening at finite temperature, and that non-perturbative effects continue to play an important role for hadrons in the chiral crossover region. This picture aligns with the expectations of chiral spin symmetry, where chiral symmetry is effectively restored but a dominant fraction of quarks remain bound~\cite{Rohrhofer:2019qwq,Rohrhofer:2019qal,Glozman:2022lda}. Although the focus of this study was correlators involving pseudo-scalar meson operators, the approach itself can in principle be generalised to other types of hadronic states of different spin, as well as to systems with non-vanishing baryon density. This could help provide a better understanding of in-medium effects in different regions of the QCD phase diagram. We leave these generalisations to future works.

\section*{Acknowledgements}
The authors acknowledge support by the Deutsche Forschungsgemeinschaft (DFG, German Research Foundation) through the Collaborative Research Center CRC-TR 211 ``Strong-interaction matter under extreme conditions'' -- Project No. 315477589-TRR 211. O.~P.~also acknowledges support by the State of Hesse within the Research Cluster ELEMENTS (Project ID 500/10.006). The authors gratefully acknowledge the Gauss Centre for Supercomputing e.V. (\texttt{www.gauss-centre.eu}) for funding this project by providing computing time through the John von Neumann Institute for Computing (NIC) on the GCS Supercomputer JUWELS~\cite{JUWELS} at the J\"{u}lich Supercomputing Centre (JSC). We also acknowledge the EuroHPC Joint Undertaking for awarding this project access to the EuroHPC supercomputer LUMI, hosted by CSC (Finland) and the LUMI consortium through a EuroHPC Extreme Scale Access call. For the calculations on these machines we were using Grid~\cite{Boyle:2015tjk,Yamaguchi:2022feu} and the Grid Python Toolkit (GPT)~\cite{GPTCode}. Part of the analysis was performed on the GPU cluster at Bielefeld University. We thank the Bielefeld HPC.NRW team for their support. \\

\noindent
All of the lattice data used in the analysis of this paper and presented in each of the figures can be found in Ref.~\cite{Data_DOI}.

\appendix

\section{Mixed action approach and mass tuning}
\label{AppendixMixedActionTuning}

Ideally, any lattice QCD action should fulfil nearest neighbour locality, be free of doublers, and have a massless Dirac operator that is chirally symmetric. However, it has been proven in even dimensions that it is not possible to simultaneously satisfy all three of these properties~\cite{Nielsen:1981hk,Nielsen:1980rz,Nielsen:1981xu}. One therefore has to weigh up the advantages and disadvantages of different lattice actions and select the one that best suits the problem at hand. For the purposes of chiral symmetry preservation, common choices are the highly-improved staggered quarks (HISQ)~\cite{Follana:2003fe,Follana:2006rc}, or a form of domain wall fermions (DWF)~\cite{Kaplan:1992bt,Kaplan:2009yg}. The HISQ action is computationally cheap compared to other lattice actions, but there only remain traces of chiral symmetry, whereas DWF actions have improved chiral characteristics, but their realisation requires the introduction of a fifth lattice dimension which makes them computationally more expensive. As a compromise, one can mix these actions to benefit from both the cheapness of the HISQs and the good chiral properties of the DWF action. \\

\noindent
For the purposes of this work we used M\"{o}bius domain wall fermion (MDWF) observables~\cite{Brower:2012vk}, and computed them on HISQ configurations. DWF actions are known to only fulfil chiral symmetry approximately when the lattice extent of the fifth dimension $L_{5}$ is finite. The reliability of this approximation can be measured by the so-called residual mass~\cite{Brower:2012vk}
\begin{align}
m_{\text{res}} = \frac{\sum_{\mathbf{x}} \langle \bar{Q}( t,\mathbf{x}) \gamma_{5} Q(t,\mathbf{x}) \bar{q}(0,\mathbf{0}) \gamma_{5} q(0, \mathbf{0})  \rangle }{ \sum_{\mathbf{x}} \langle \bar{q}(t,\mathbf{x}) \gamma_{5} q(t, \mathbf{x}) \bar{q}(0, \mathbf{0}) \gamma_{5} q(0, \mathbf{0})  \rangle },
 \end{align}
where $Q(t,\mathbf{x})$ is a quark field in the midpoint of the fifth dimension, and $q\left( t,\mathbf{x} \right)$ is a quark field bound to the domain wall. $m_{\text{res}}$ depends on the DWF parameters, which in the specific case of MDWFs are $L_{5}$, the domain wall height $M_{5}$, and $b_{5}, c_{5}$, which are related to the lattice spacing of the fifth dimension $a_{5}$ via: $a_{5} = b_{5}-c_{5}$. In the limit $L_{5}\rightarrow \infty$ the residual mass vanishes, and hence chiral symmetry is restored. \\

\noindent
A drawback of the mixed action approach is that an additional mass term $\Delta m_{\text{mix}}$~\cite{Bar:2005tu} appears which causes the bare sea and valence quark masses to differ, thus requiring the valence quark mass to be tuned. We performed this tuning on the lowest-temperature configurations detailed in Tables~\ref{tabconf_ls} and~\ref{tabconf_ss} in two steps:

\begin{enumerate}

\item Adjust the MDWF parameters to reduce $m_{\text{res}}$ such that its value is to within 10\% of the light valence bare quark mass $m_{l,\text{val}}$. 
    
\item Vary $m_{q,\text{val}}$ for $q \in \lbrace l,s \rbrace$ until the pseudo-scalar meson ground state mass $m_{q\bar{q}}$ is consistent with the corresponding physical neutrally charged particle mass, as listed in the PDG~\cite{Workman:2022ynf}.  
\end{enumerate}

\begin{table}[ht!]
\center
\small
\begin{tabular}{|c|c|c|c|c|c|c|c|} 
\hline
\rule{0pt}{3ex}
\!\!\!\! $T \, [\text{MeV}]$ \!\!\!\! & $a \, [\text{fm}]$  &  $\beta$  & $N_{s}^{3} \times N_{\tau}$ & $am_{l,\text{val}}$ & $am_{s,\text{val}}$ & $am_{l,\text{sea}}$ & $am_{s,\text{sea}}$    \\[0.5ex]
\hhline{|=|=|=|=|=|=|=|=|}
\!\!\!\! 36.4  \!\!\!\!  & 0.085 & 7.010  & $64^{3} \times 64$ & $0.0010$  & $0.0378$ & $0.0013$  & $0.0357$   \\ 
 \hline
\!\!\!\! 43.1  \!\!\!\!  & 0.072 & 7.188  & $64^{3} \times 64$ & $0.0008$  & $0.0297$ & $0.0011$  & $0.0306$  \\ 
 \hline
\end{tabular}
\caption{Bare sea quark masses $m_{q,\text{sea}}$ of the HISQ configurations and the corresponding tuned bare valence quark masses $m_{q,\text{val}}$ for $q \in \lbrace l, s \rbrace$. For both values of $\beta$ the MDWF parameters were set to: $L_{5} = 16$, $M_{5} = 1.3$, $b_{5} = 1.75$, and $c_{5} = 1 - b_{5}$.}
\label{tabtuning}
\end{table}

\begin{figure}[ht!]
\centering
\includegraphics[width=0.4\textwidth]{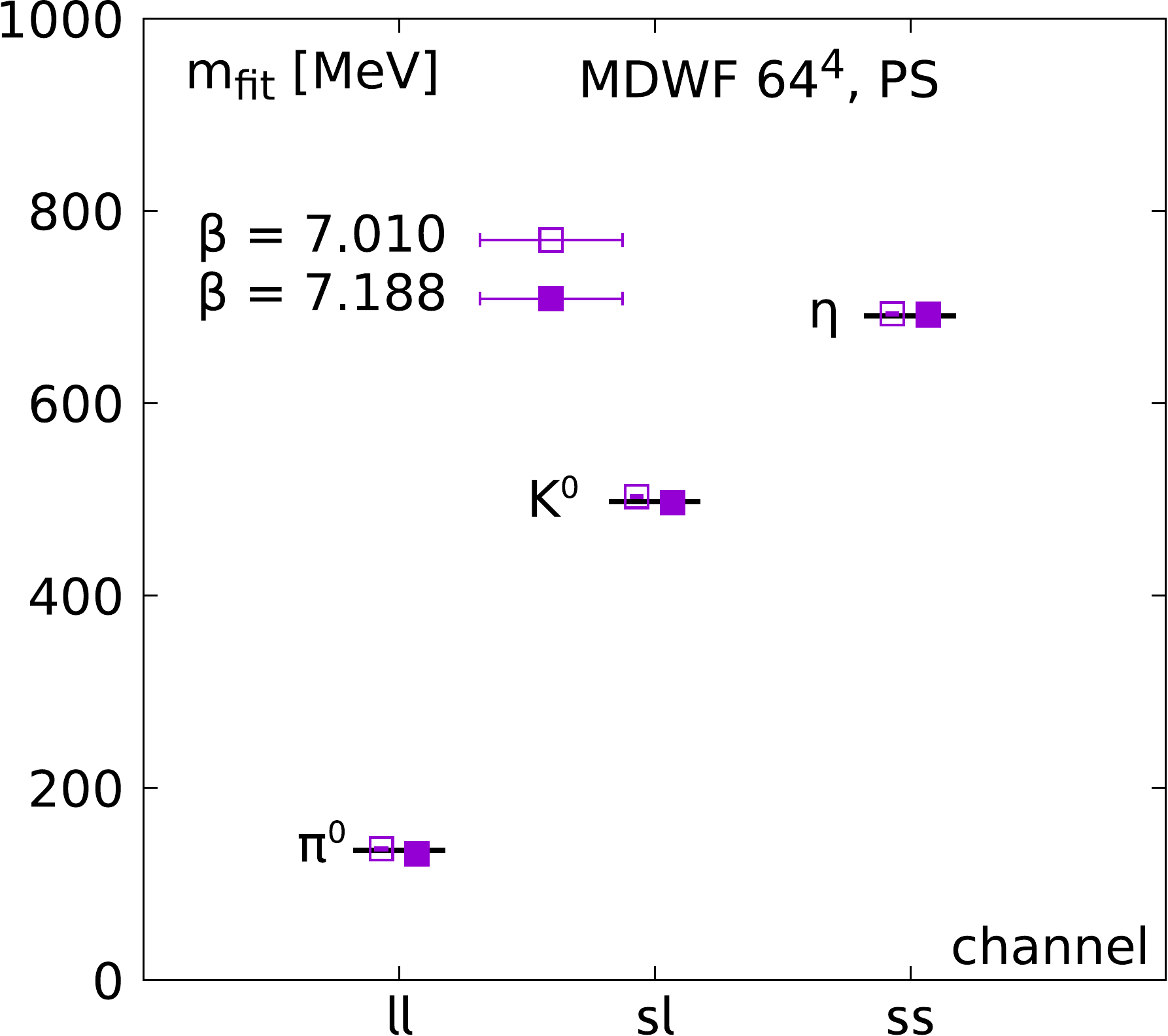}
\caption{Pseudo-scalar meson ground state masses calculated with the tuned bare quark masses in Tab.~\ref{tabtuning}, and the corresponding physical neutrally charged particle masses from the PDG~\cite{Workman:2022ynf}. Since there is no pure pseudo-scalar $s\bar{s}$ bound state, we used the value $m_{\eta}=\sqrt{2(m_{K^{0}})^{2} - (m_{\pi^{0}})^{2}}$ in order to tune the strange bare quark mass.}
\label{fig:Spectrumplot}
\end{figure}

\noindent
The parameter values from this procedure are listed in Tab.~\ref{tabtuning}, and in Fig.~\ref{fig:Spectrumplot} is shown the comparison between the physical ground state masses and those extracted using the tuned parameters. \\

\clearpage

\section{Stability analysis results}
\label{stabil_analysis}

\begin{figure}[h!]
\centering
\includegraphics[width=0.41\textwidth]{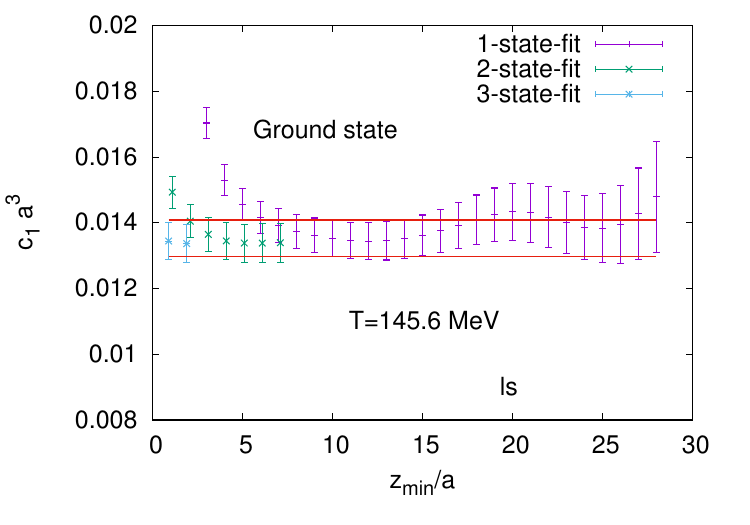}
\includegraphics[width=0.41\textwidth]{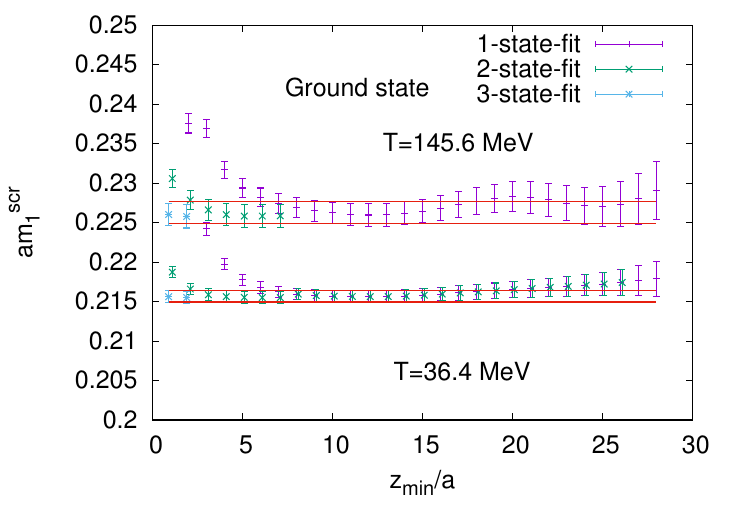}
\includegraphics[width=0.41\textwidth]{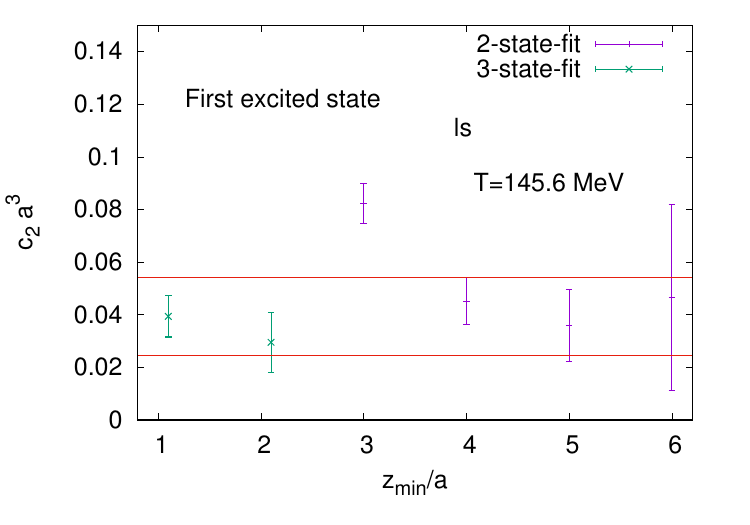}
\includegraphics[width=0.41\textwidth]{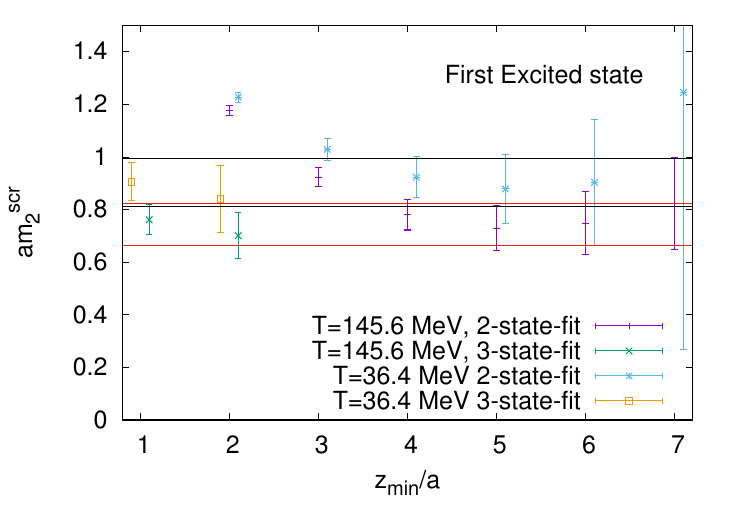}
\vspace{-3mm}
\caption{$\bar{l} \gamma_{5} s$ stability analysis at $T=145.6 \, \text{MeV}$ for $\left\{c_{i},m_{i}^{\text{scr}}\right\}$. See Sec.~\ref{lattice_setup} for the fit parameter definitions.}
\label{ls_145_st}
\vspace{5mm}
\centering
\includegraphics[width=0.41\textwidth]{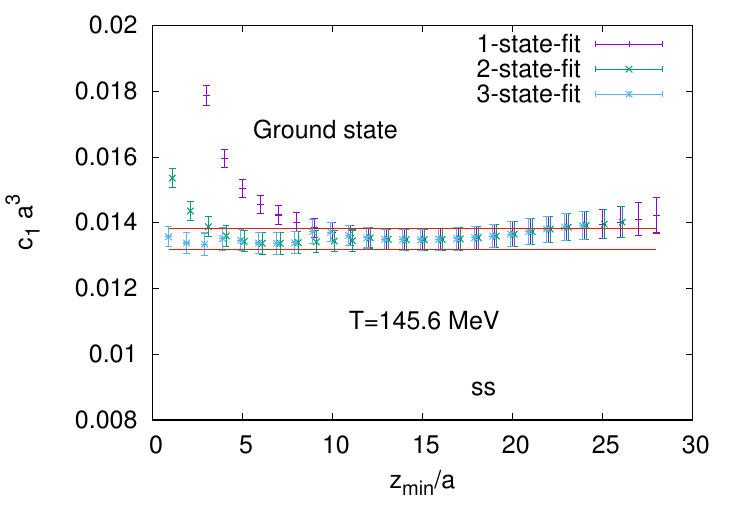}
\includegraphics[width=0.41\textwidth]{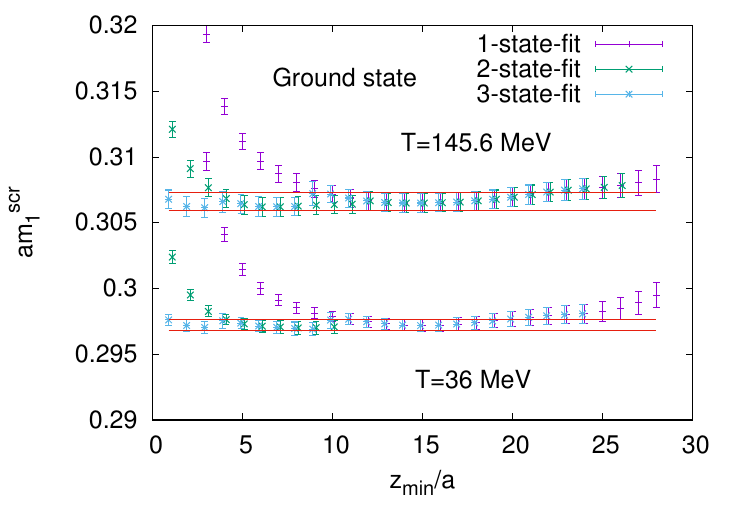}
\includegraphics[width=0.41\textwidth]{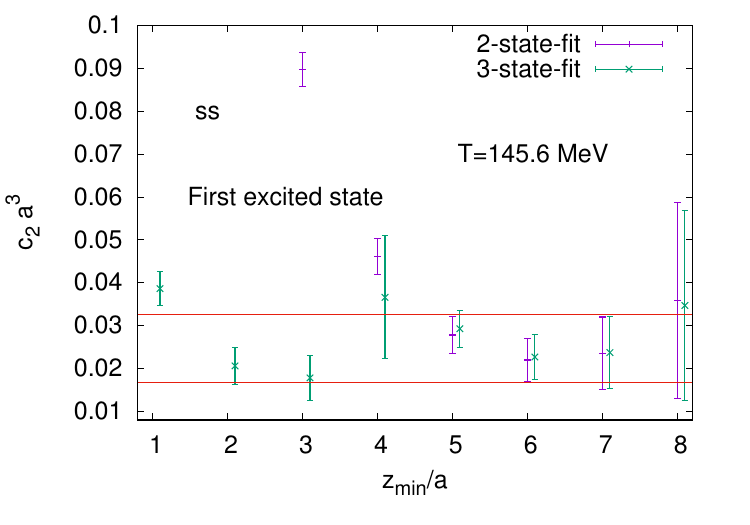}
\includegraphics[width=0.41\textwidth]{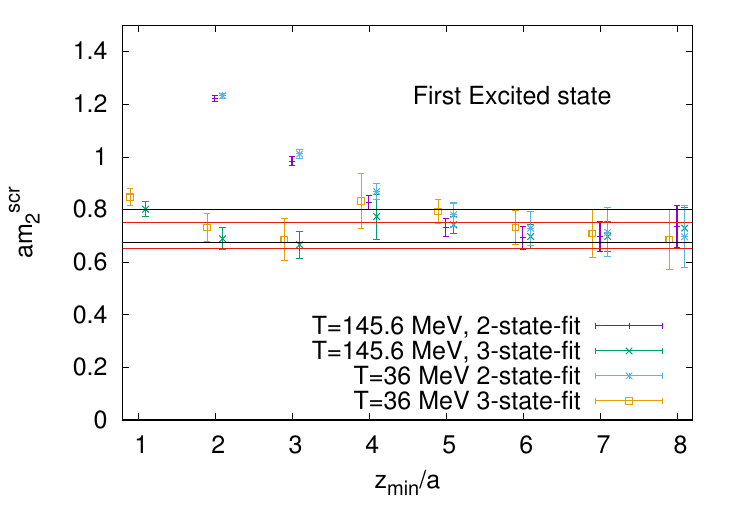}
\vspace{-3mm}
\caption{$\bar{s} \gamma_{5} s$ stability analysis at $T=145.6 \, \text{MeV}$ for $\left\{c_{i},m_{i}^{\text{scr}}\right\}$. See Sec.~\ref{lattice_setup} for the fit parameter definitions.}
\label{ss_145_st}
\end{figure}

\begin{figure}[h!]
\centering
\includegraphics[width=0.41\textwidth]{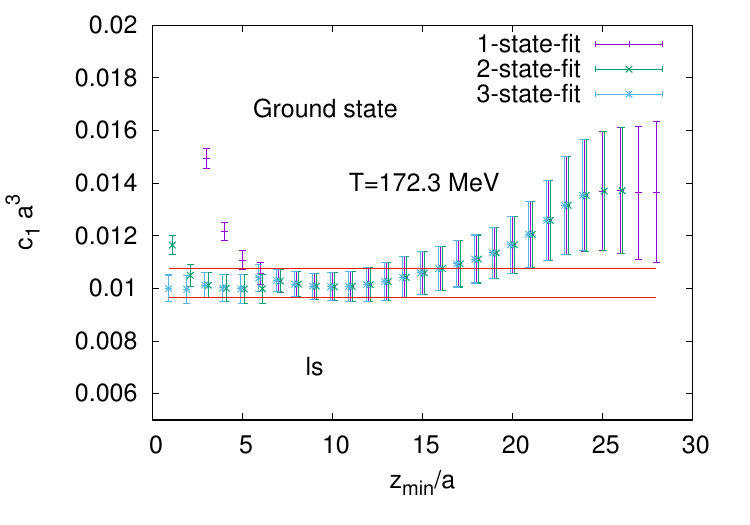}
\includegraphics[width=0.41\textwidth]{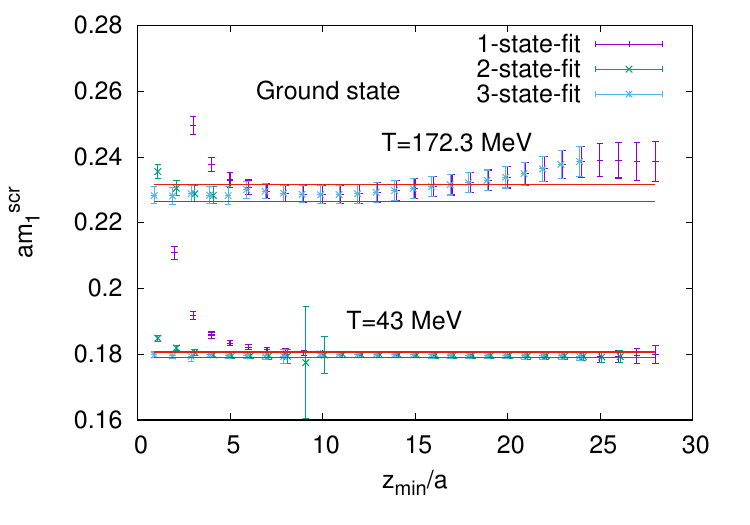}
\includegraphics[width=0.41\textwidth]{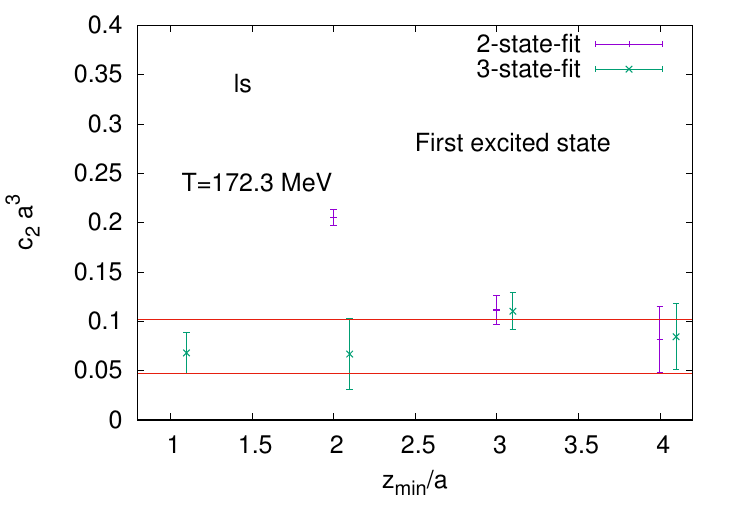}
\includegraphics[width=0.41\textwidth]{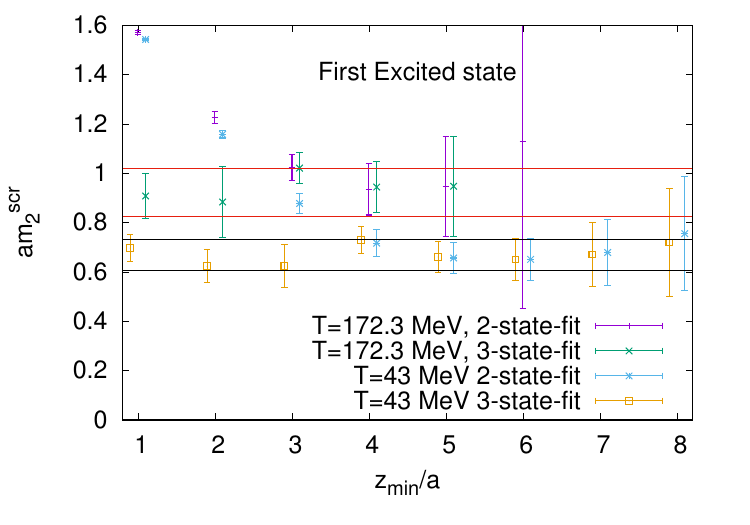}
\vspace{-3mm}
\caption{$\bar{l} \gamma_{5} s$ stability analysis at $T=172.3 \, \text{MeV}$ for $\left\{c_{i},m_{i}^{\text{scr}}\right\}$. See Sec.~\ref{lattice_setup} for the fit parameter definitions.}
\label{ls_172_st}
\vspace{13mm}
\centering
\includegraphics[width=0.41\textwidth]{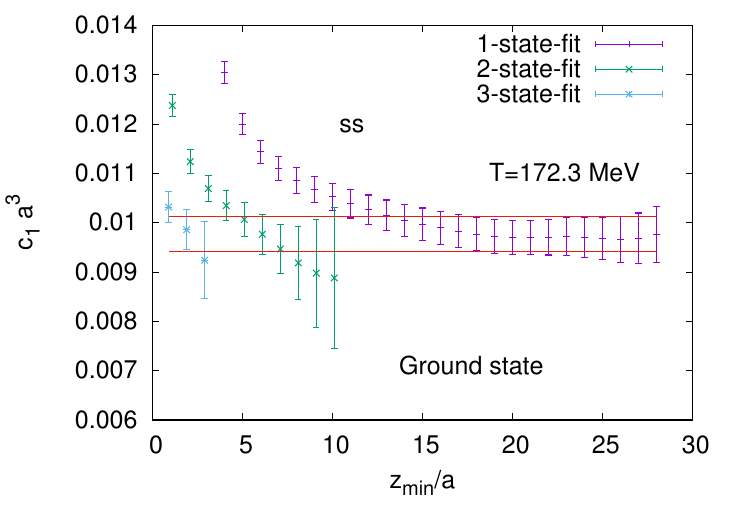}
\includegraphics[width=0.41\textwidth]{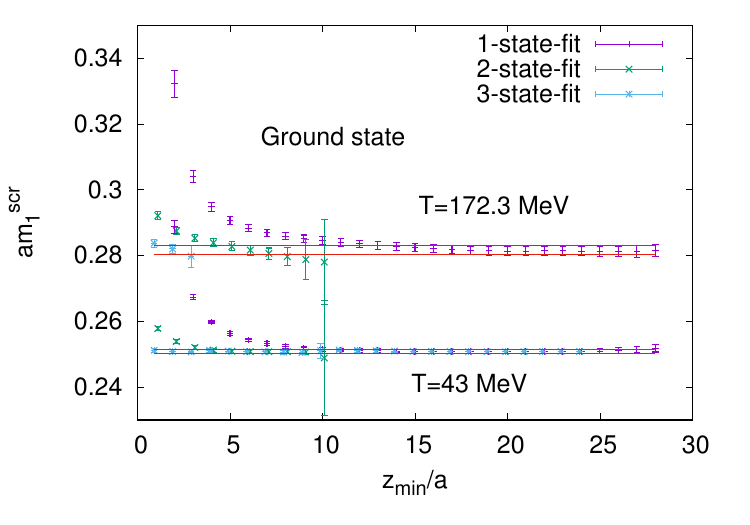}
\includegraphics[width=0.41\textwidth]{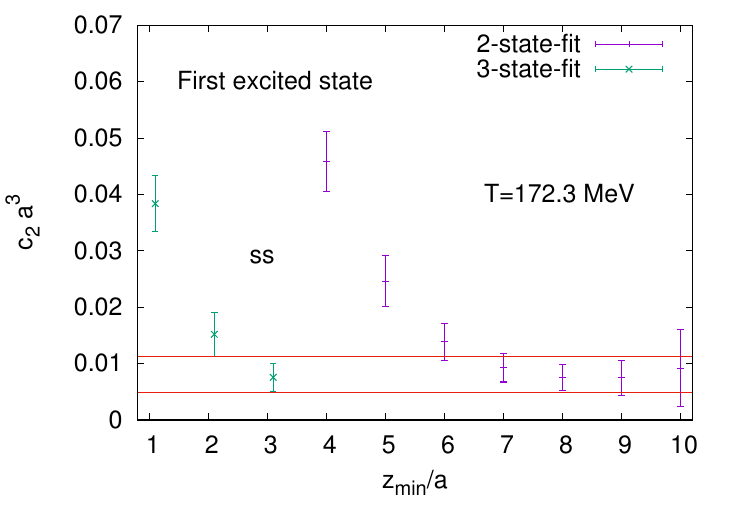}
\includegraphics[width=0.41\textwidth]{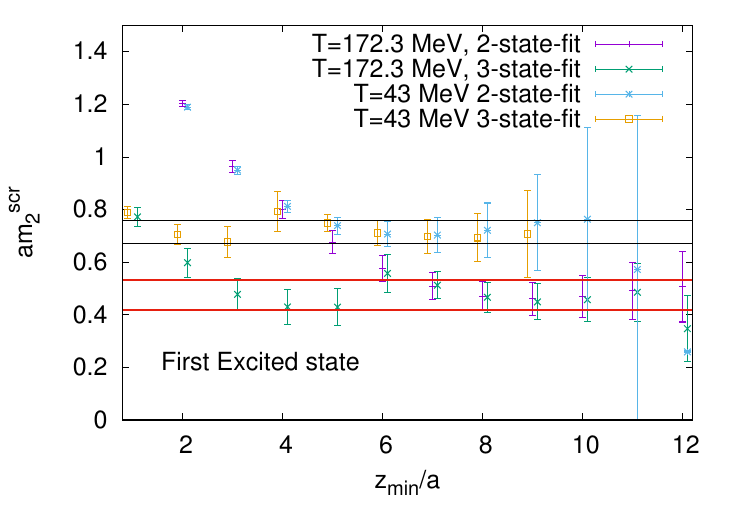}
\vspace{-3mm}
\caption{$\bar{s} \gamma_{5} s$ stability analysis at $T=172.3 \, \text{MeV}$ for $\left\{c_{i},m_{i}^{\text{scr}}\right\}$. See Sec.~\ref{lattice_setup} for the fit parameter definitions.}
\label{ss_172_st}
\end{figure}

\clearpage

\section{Prediction quality}
\label{pred_quality}

To provide a quantitative measure of the agreement between the temporal correlator data $\widetilde{C}_{\text{PS}}^{\text{data}}$ and predictions $\widetilde{C}_{\text{PS}}^{\text{pred}}$ we compute the mean absolute percentage error (MAPE), defined as
\begin{align}
\text{MAPE} = 100 \% \times \frac{1}{N_{\tau}/2 - n_{\tau}^{*} +1}\sum_{n_{\tau}=n_{\tau}^{*}}^{N_{\tau}/2} \left| \frac{ \widetilde{C}_{\text{PS}}^{\text{latt}}(\tau = n_{\tau}a,\vec{p})-\widetilde{C}_{\text{PS}}^{\text{pred}}(\tau = n_{\tau}a,\vec{p})}{\widetilde{C}_{\text{PS}}^{\text{latt}}(\tau = n_{\tau}a,\vec{p})} \right|,
\end{align}
where $n_{\tau}^{*}$ is the first $\tau$ point above the dashed line of Figs~\ref{ls_145_specT},~\ref{ls_172_specT},~\ref{ss_145_specT}-\ref{ss_172_specT} where the prediction is applicable. Furthermore, we define $\Delta_{\text{stat}}$ to be the average of the statistical error bars on the lattice data, and $\Delta_{\text{theor}}$ the average of the bootstrap-computed error band on the theoretical prediction. Details of the error band computation can be found in Sec.~\ref{lattice_setup}. The values of MAPE, $\Delta_{\text{stat}}$, and $\Delta_{\text{theor}}$ are displayed in Table~\ref{MAPE} for each channel and spatial momenta. If the MAPE is comparable or smaller than the combined lattice and theoretical errors, namely: $\text{MAPE} \lesssim \Delta_{\text{stat}} + \Delta_{\text{theor}}$, the prediction provides a good description of the data.

\begin{table}[h!]
\center
\footnotesize
\renewcommand{\arraystretch}{1.151}
\begin{tabular}{|c|c|c|c|c|c|} 
\hline
\rule{0pt}{3ex}
Channel & \!\!\!\! $T \, [\text{MeV}]$ \!\!\!\! & $p \, [\text{MeV}]$  &  MAPE [\%]  & $\Delta_{\text{stat}}$ [\%] & $\Delta_{\text{theor}}$ [\%]   \\[0.5ex]
\hhline{|=|=|=|=|=|=|}
$\bar{l} \gamma_{5} s$ & 145.6	& 0		& 8.4	& 2.9 & 8.0\\
\hline
$\bar{l} \gamma_{5} s$ & 145.6	& 229	& 8.9	& 2.9 & 9.0\\
\hline
$\bar{l} \gamma_{5} s$ & 145.6	& 457	& 11.1	& 3.0 & 11.5\\
\hline
$\bar{l} \gamma_{5} s$ & 145.6	& 686	& 15.8	& 3.4 & 15.6 \\
\hline
$\bar{l} \gamma_{5} s$ & 145.6	& 914	& 22.6	& 4.3 & 20.5 \\
\hhline{|=|=|=|=|=|=|}
$\bar{l} \gamma_{5} s$ & 172.3	& 0		& 10.8	& 3.0 & 6.7\\
\hline
$\bar{l} \gamma_{5} s$ & 172.3	& 271	& 12.2	& 2.9 & 7.3\\
\hline
$\bar{l} \gamma_{5} s$ & 172.3	& 541	& 15.3	& 3.0 & 9.4\\
\hline
$\bar{l} \gamma_{5} s$ & 172.3	& 812	& 20.8	& 3.3 & 12.9 \\
\hline
$\bar{l} \gamma_{5} s$ & 172.3	& 1082	& 30.9	& 4.1 & 17.8 \\
\hhline{|=|=|=|=|=|=|}
$\bar{s} \gamma_{5} s$ & 145.6 & 0		& 3.5	& 1.7 & 9.2\\
\hline
$\bar{s} \gamma_{5} s$ & 145.6 & 229		& 3.7	& 1.7 & 9.9\\
\hline
$\bar{s} \gamma_{5} s$ & 145.6 & 457		& 4.4	& 1.8 & 11.9\\
\hline
$\bar{s} \gamma_{5} s$ & 145.6 & 686		& 6.3	& 2.0 & 15.0 \\
\hline
$\bar{s} \gamma_{5} s$ & 145.6 & 914		& 9.7	& 2.4 & 18.4\\
\hhline{|=|=|=|=|=|=|}
$\bar{s} \gamma_{5} s$ & 172.3 & 0		& 1.4	& 2.2 & 17.0\\
\hline
$\bar{s} \gamma_{5} s$ & 172.3 & 271		& 1.5	& 2.3 & 17.7\\
\hline
$\bar{s} \gamma_{5} s$ & 172.3 & 541 	& 1.6	& 2.4 & 19.4\\
\hline
$\bar{s} \gamma_{5} s$ & 172.3 & 812		& 3.1	& 2.6 & 21.0 \\
\hline
$\bar{s} \gamma_{5} s$ & 172.3 & 1082	& 7.5	& 2.9 & 22.1\\
\hhline{|=|=|=|=|=|=|}
$\bar{s} \gamma_{5} s$ & 172.3 & 0		& 10.3	& 2.2 & 3.9\\
\hline
$\bar{s} \gamma_{5} s$ & 172.3 & 271		& 11.4	& 2.3 & 4.1\\
\hline
$\bar{s} \gamma_{5} s$ & 172.3 & 541		& 14.6	& 2.4 & 4.5\\
\hline
$\bar{s} \gamma_{5} s$ & 172.3 & 812		& 19.3	& 2.6 & 5.3 \\
\hline
$\bar{s} \gamma_{5} s$ & 172.3 & 1082	& 24.7	& 2.9 & 6.2\\
\hline
\end{tabular}
\caption{The mean absolute percentage error (MAPE) of the $\widetilde{C}_{\text{PS}}(\tau,\vec{p})$ predictions, average statistical lattice error, $\Delta_{\text{stat}}$, and theoretical uncertainty, $\Delta_{\text{theor}}$, across the prediction range. In descending order these values correspond to the data and predictions in Figs.~\ref{ls_145_specT},~\ref{ls_172_specT},~\ref{ss_145_specT},~\ref{ss_172_specT2}, and~\ref{ss_172_specT}, respectively.}
\label{MAPE}
\end{table}

\newpage

\section{Perturbative comparison}
\label{pert_pred}

\begin{figure}[h!] 
\centering 
\includegraphics[width=0.49\textwidth]{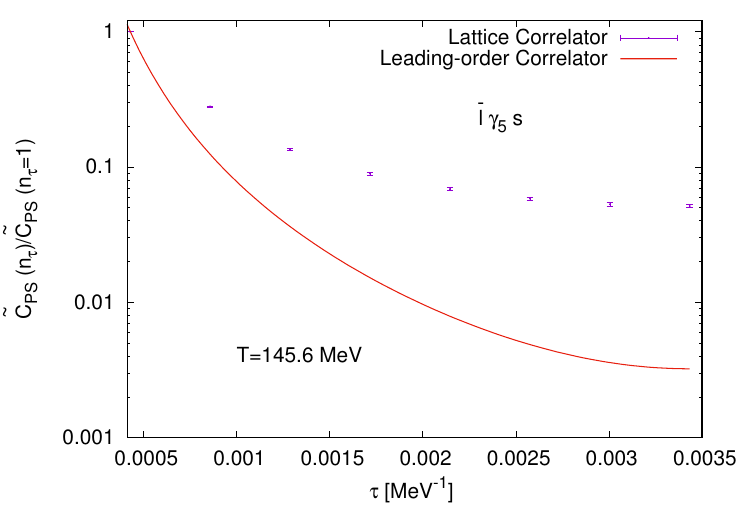}
\includegraphics[width=0.49\textwidth]{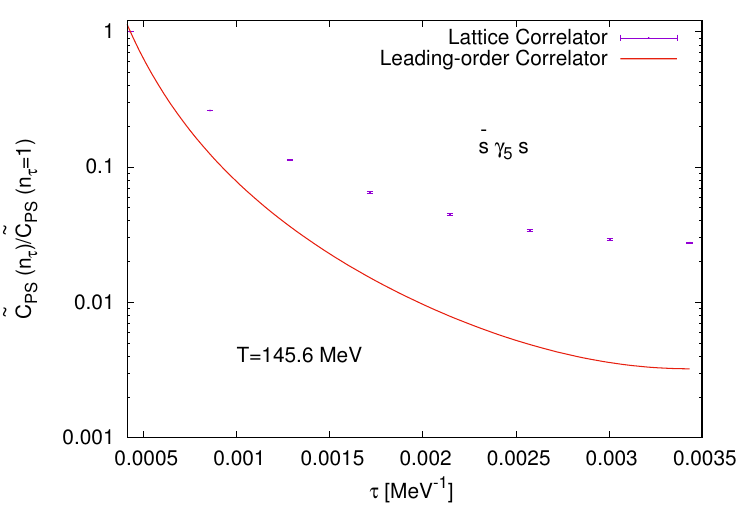}
\includegraphics[width=0.49\textwidth]{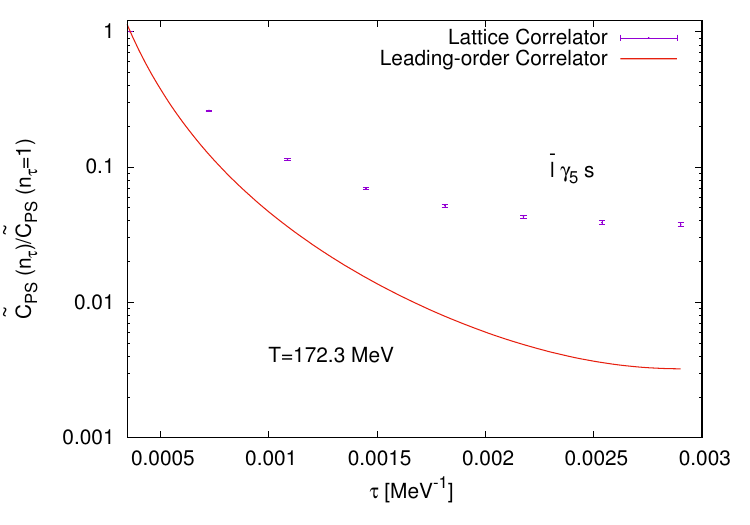}
\includegraphics[width=0.49\textwidth]{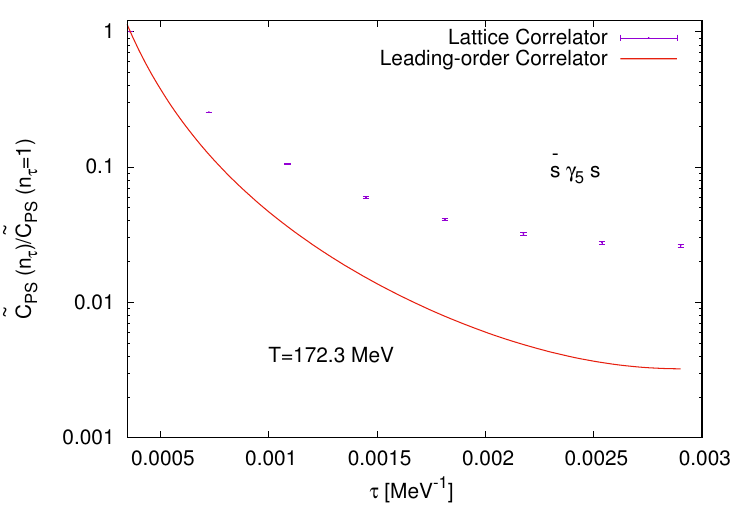}
\caption{Comparison of the leading-order perturbative prediction for $\widetilde{C}_{\text{PS}}(\tau,\vec{p}=0)/\widetilde{C}_{\text{PS}}(n_{\tau}=1,\vec{p}=0)$ with the lattice data in each channel at $T=145.6 \, \text{MeV}$ and $172.3 \, \text{MeV}$.}
\label{Tcorr_pert}
\end{figure}

In Fig.~\ref{Tcorr_pert} we compare the leading-order perturbative predictions for the temporal correlator at zero momentum with the lattice data at $T=145.6 \, \text{MeV}$ and $172.3 \, \text{MeV}$ in both channels. To remove unknown renormalisation constants we perform this comparison for the unit-normalised correlator at $n_{\tau}=1$. Since the quark masses are small relative to those of the meson ground states, the prediction is computed for massless quarks using the spectral function derived in Ref.~\cite{Aarts:2005hg}. From Fig.~\ref{Tcorr_pert} one can see that the perturbative predictions significantly underestimate the data for large $\tau$. This demonstrates that these predictions only affect the large-energy behaviour of the spectral function, which dominates the small $\tau$ structure of $\widetilde{C}_{\text{PS}}(\tau,\vec{p}=0)$. It should be noted that the perturbative predictions displayed in Fig.~\ref{Tcorr_pert} are continuum results, and therefore do not take lattice discretisation effects into account. However, such effects are generally small in the low-energy regime~\cite{Aarts:2005hg}, and would not significantly alter the large-$\tau$ predictions. These results demonstrate that the perturbative contributions to $\rho_{\text{PS}}(\omega,\vec{p})$ are highly sub-dominant at low energies, and hence the spectral components analysed in Sec.~\ref{PS_analysis} are of a non-perturbative nature. One expects that this sub-dominance will change at higher temperatures due to the increasing importance of non-thermoparticle-like components, as outlined in Sec.~\ref{ss_channel}.

\vspace{5mm}

\bibliographystyle{JHEP}

\bibliography{PS_refs}

\providecommand{\href}[2]{#2}\begingroup\raggedright\begin{thebibliography}{10}

\bibitem{Detar:1987kae}
C.~E. Detar and J.~B. Kogut, \emph{{The Hadronic Spectrum of the Quark
  Plasma}}, \href{https://doi.org/10.1103/PhysRevLett.59.399}{\emph{Phys. Rev.
  Lett.} {\bfseries 59} (1987) 399}.

\bibitem{Born:1991zz}
{\scshape MT(c)} collaboration, \emph{{Hadronic correlation functions in the
  QCD plasma phase}},
  \href{https://doi.org/10.1103/PhysRevLett.67.302}{\emph{Phys. Rev. Lett.}
  {\bfseries 67} (1991) 302}.

\bibitem{Florkowski:1993bq}
W.~Florkowski and B.~L. Friman, \emph{{Spatial dependence of the finite
  temperature meson correlation function}},
  \href{https://doi.org/10.1007/BF01289794}{\emph{Z. Phys. A} {\bfseries 347}
  (1994) 271}.

\bibitem{Kogut:1998rh}
J.~B. Kogut, J.~F. Lagae and D.~K. Sinclair, \emph{{Topology, fermionic zero
  modes and flavor singlet correlators in finite temperature QCD}},
  \href{https://doi.org/10.1103/PhysRevD.58.054504}{\emph{Phys. Rev. D}
  {\bfseries 58} (1998) 054504}
  [\href{https://arxiv.org/abs/hep-lat/9801020}{{\ttfamily hep-lat/9801020}}].

\bibitem{Aarts:2005hg}
G.~Aarts and J.~M. Martinez~Resco, \emph{{Continuum and lattice meson spectral
  functions at nonzero momentum and high temperature}},
  \href{https://doi.org/10.1016/j.nuclphysb.2005.08.012}{\emph{Nucl. Phys. B}
  {\bfseries 726} (2005) 93}
  [\href{https://arxiv.org/abs/hep-lat/0507004}{{\ttfamily hep-lat/0507004}}].

\bibitem{Wetzorke:2001dk}
I.~Wetzorke, F.~Karsch, E.~Laermann, P.~Petreczky and S.~Stickan, \emph{{Meson
  spectral functions at finite temperature}},
  \href{https://doi.org/10.1016/S0920-5632(01)01763-7}{\emph{Nucl. Phys. B
  Proc. Suppl.} {\bfseries 106} (2002) 510}
  [\href{https://arxiv.org/abs/hep-lat/0110132}{{\ttfamily hep-lat/0110132}}].

\bibitem{Karsch:2003jg}
F.~Karsch and E.~Laermann, \emph{{Thermodynamics and in medium hadron
  properties from lattice QCD}},
  \href{https://arxiv.org/abs/hep-lat/0305025}{{\ttfamily hep-lat/0305025}}.

\bibitem{Petreczky:2003iz}
P.~Petreczky, \emph{{Lattice calculations of meson correlators and spectral
  functions at finite temperature}},
  \href{https://doi.org/10.1088/0954-3899/30/1/051}{\emph{J. Phys. G}
  {\bfseries 30} (2004) S431}
  [\href{https://arxiv.org/abs/hep-ph/0305189}{{\ttfamily hep-ph/0305189}}].

\bibitem{Asakawa:2003re}
M.~Asakawa and T.~Hatsuda, \emph{{J / psi and eta(c) in the deconfined plasma
  from lattice QCD}},
  \href{https://doi.org/10.1103/PhysRevLett.92.012001}{\emph{Phys. Rev. Lett.}
  {\bfseries 92} (2004) 012001}
  [\href{https://arxiv.org/abs/hep-lat/0308034}{{\ttfamily hep-lat/0308034}}].

\bibitem{Aarts:2006em}
{\scshape UKQCD} collaboration, \emph{{Meson spectral functions with chirally
  symmetric lattice fermions}},
  \href{https://doi.org/10.1088/1126-6708/2007/02/062}{\emph{JHEP} {\bfseries
  02} (2007) 062} [\href{https://arxiv.org/abs/hep-lat/0612007}{{\ttfamily
  hep-lat/0612007}}].

\bibitem{Ding:2012sp}
H.~T. Ding, A.~Francis, O.~Kaczmarek, F.~Karsch, H.~Satz and W.~Soeldner,
  \emph{{Charmonium properties in hot quenched lattice QCD}},
  \href{https://doi.org/10.1103/PhysRevD.86.014509}{\emph{Phys. Rev. D}
  {\bfseries 86} (2012) 014509}
  [\href{https://arxiv.org/abs/1204.4945}{{\ttfamily 1204.4945}}].

\bibitem{Burnier:2015tda}
Y.~Burnier, O.~Kaczmarek and A.~Rothkopf, \emph{{Quarkonium at finite
  temperature: Towards realistic phenomenology from first principles}},
  \href{https://doi.org/10.1007/JHEP12(2015)101}{\emph{JHEP} {\bfseries 12}
  (2015) 101} [\href{https://arxiv.org/abs/1509.07366}{{\ttfamily
  1509.07366}}].

\bibitem{Mukherjee:2015mxc}
S.~Mukherjee, P.~Petreczky and S.~Sharma, \emph{{Charm degrees of freedom in
  the quark gluon plasma}},
  \href{https://doi.org/10.1103/PhysRevD.93.014502}{\emph{Phys. Rev. D}
  {\bfseries 93} (2016) 014502}
  [\href{https://arxiv.org/abs/1509.08887}{{\ttfamily 1509.08887}}].

\bibitem{Meyer:2017ydp}
H.~Meyer, \emph{{Lattice QCD, Spectral Functions and Transport Coefficients}},
  \href{https://doi.org/10.22323/1.281.0364}{\emph{PoS} {\bfseries INPC2016}
  (2017) 364}.

\bibitem{Rothkopf:2019ipj}
A.~Rothkopf, \emph{{Heavy Quarkonium in Extreme Conditions}},
  \href{https://doi.org/10.1016/j.physrep.2020.02.006}{\emph{Phys. Rept.}
  {\bfseries 858} (2020) 1} [\href{https://arxiv.org/abs/1912.02253}{{\ttfamily
  1912.02253}}].

\bibitem{Kapusta:2006pm}
J.~I. Kapusta and C.~Gale, \emph{{Finite-temperature Field Theory: Principles
  and applications}}, Cambridge Monographs on Mathematical Physics. Cambridge
  University Press, 2011.

\bibitem{Bellac:2011kqa}
M.~L. Bellac, \emph{{Thermal Field Theory}}, Cambridge Monographs on
  Mathematical Physics. Cambridge University Press, 3, 2011.

\bibitem{Asakawa:2000tr}
M.~Asakawa, T.~Hatsuda and Y.~Nakahara, \emph{{Maximum entropy analysis of the
  spectral functions in lattice QCD}},
  \href{https://doi.org/10.1016/S0146-6410(01)00150-8}{\emph{Prog. Part. Nucl.
  Phys.} {\bfseries 46} (2001) 459}
  [\href{https://arxiv.org/abs/hep-lat/0011040}{{\ttfamily hep-lat/0011040}}].

\bibitem{Meyer:2011gj}
H.~B. Meyer, \emph{{Transport Properties of the Quark-Gluon Plasma: A Lattice
  QCD Perspective}},
  \href{https://doi.org/10.1140/epja/i2011-11086-3}{\emph{Eur. Phys. J. A}
  {\bfseries 47} (2011) 86} [\href{https://arxiv.org/abs/1104.3708}{{\ttfamily
  1104.3708}}].

\bibitem{Laermann:2001vg}
E.~Laermann and P.~Schmidt, \emph{{Meson screening masses at high temperature
  in quenched QCD with improved Wilson quarks}},
  \href{https://doi.org/10.1007/s100520100682}{\emph{Eur. Phys. J. C}
  {\bfseries 20} (2001) 541}
  [\href{https://arxiv.org/abs/hep-lat/0103037}{{\ttfamily hep-lat/0103037}}].

\bibitem{Wissel:2005pb}
S.~Wissel, E.~Laermann, S.~Shcheredin, S.~Datta and F.~Karsch, \emph{{Meson
  correlation functions at high temperatures}},
  \href{https://doi.org/10.22323/1.020.0164}{\emph{PoS} {\bfseries LAT2005}
  (2006) 164} [\href{https://arxiv.org/abs/hep-lat/0510031}{{\ttfamily
  hep-lat/0510031}}].

\bibitem{Cheng:2010fe}
M.~Cheng et~al., \emph{{Meson screening masses from lattice QCD with two light
  and the strange quark}},
  \href{https://doi.org/10.1140/epjc/s10052-011-1564-y}{\emph{Eur. Phys. J. C}
  {\bfseries 71} (2011) 1564}
  [\href{https://arxiv.org/abs/1010.1216}{{\ttfamily 1010.1216}}].

\bibitem{Banerjee:2011yd}
D.~Banerjee, R.~V. Gavai and S.~Gupta, \emph{{Quasi-static probes of the QCD
  plasma}}, \href{https://doi.org/10.1103/PhysRevD.83.074510}{\emph{Phys. Rev.
  D} {\bfseries 83} (2011) 074510}
  [\href{https://arxiv.org/abs/1102.4465}{{\ttfamily 1102.4465}}].

\bibitem{Karsch:2012na}
F.~Karsch, E.~Laermann, S.~Mukherjee and P.~Petreczky, \emph{{Signatures of
  charmonium modification in spatial correlation functions}},
  \href{https://doi.org/10.1103/PhysRevD.85.114501}{\emph{Phys. Rev. D}
  {\bfseries 85} (2012) 114501}
  [\href{https://arxiv.org/abs/1203.3770}{{\ttfamily 1203.3770}}].

\bibitem{Brandt:2014uda}
B.~B. Brandt, A.~Francis, M.~Laine and H.~B. Meyer, \emph{{A relation between
  screening masses and real-time rates}},
  \href{https://doi.org/10.1007/JHEP05(2014)117}{\emph{JHEP} {\bfseries 05}
  (2014) 117} [\href{https://arxiv.org/abs/1404.2404}{{\ttfamily 1404.2404}}].

\bibitem{Bazavov:2014cta}
A.~Bazavov, F.~Karsch, Y.~Maezawa, S.~Mukherjee and P.~Petreczky,
  \emph{{In-medium modifications of open and hidden strange-charm mesons from
  spatial correlation functions}},
  \href{https://doi.org/10.1103/PhysRevD.91.054503}{\emph{Phys. Rev. D}
  {\bfseries 91} (2015) 054503}
  [\href{https://arxiv.org/abs/1411.3018}{{\ttfamily 1411.3018}}].

\bibitem{Bazavov:2019www}
A.~Bazavov et~al., \emph{{Meson screening masses in (2+1)-flavor QCD}},
  \href{https://doi.org/10.1103/PhysRevD.100.094510}{\emph{Phys. Rev. D}
  {\bfseries 100} (2019) 094510}
  [\href{https://arxiv.org/abs/1908.09552}{{\ttfamily 1908.09552}}].

\bibitem{DallaBrida:2021ddx}
M.~Dalla~Brida, L.~Giusti, T.~Harris, D.~Laudicina and M.~Pepe,
  \emph{{Non-perturbative thermal QCD at all temperatures: the case of mesonic
  screening masses}},
  \href{https://doi.org/10.1007/JHEP04(2022)034}{\emph{JHEP} {\bfseries 04}
  (2022) 034} [\href{https://arxiv.org/abs/2112.05427}{{\ttfamily
  2112.05427}}].

\bibitem{Lowdon:2022xcl}
P.~Lowdon and O.~Philipsen, \emph{{Pion spectral properties above the chiral
  crossover of QCD}},
  \href{https://doi.org/10.1007/JHEP10(2022)161}{\emph{JHEP} {\bfseries 10}
  (2022) 161} [\href{https://arxiv.org/abs/2207.14718}{{\ttfamily
  2207.14718}}].

\bibitem{Bros:1992ey}
J.~Bros and D.~Buchholz, \emph{{Particles and propagators in relativistic
  thermo field theory}}, \href{https://doi.org/10.1007/BF01565114}{\emph{Z.
  Phys. C} {\bfseries 55} (1992) 509}.

\bibitem{Buchholz:1993kp}
D.~Buchholz, \emph{{On the manifestations of particles}},  in
  \emph{{International Conference on Mathematical Physics Towards the 21st
  Century}}, 11, 1993, \href{https://arxiv.org/abs/hep-th/9511023}{{\ttfamily
  hep-th/9511023}}.

\bibitem{Bros:1995he}
J.~Bros and D.~Buchholz, \emph{{Relativistic KMS condition and Kallen-Lehmann
  type representations of thermal propagators}},  in \emph{{4th Workshop on
  Thermal Field Theories and Their Applications}}, pp.~103--110, 8, 1995,
  \href{https://arxiv.org/abs/hep-th/9511022}{{\ttfamily hep-th/9511022}}.

\bibitem{Bros:1996mw}
J.~Bros and D.~Buchholz, \emph{{Axiomatic analyticity properties and
  representations of particles in thermal quantum field theory}}, {\emph{Ann.
  Inst. H. Poincare Phys. Theor.} {\bfseries 64} (1996) 495}
  [\href{https://arxiv.org/abs/hep-th/9606046}{{\ttfamily hep-th/9606046}}].

\bibitem{Bros:2001zs}
J.~Bros and D.~Buchholz, \emph{{Asymptotic dynamics of thermal quantum
  fields}}, \href{https://doi.org/10.1016/S0550-3213(02)00059-7}{\emph{Nucl.
  Phys. B} {\bfseries 627} (2002) 289}
  [\href{https://arxiv.org/abs/hep-ph/0109136}{{\ttfamily hep-ph/0109136}}].

\bibitem{Bros:1998ua}
J.~Bros and D.~Buchholz, \emph{{Towards a relativistic KMS condition}},
  \href{https://doi.org/10.1016/0550-3213(94)00298-3}{\emph{Nucl. Phys. B}
  {\bfseries 429} (1994) 291}
  [\href{https://arxiv.org/abs/hep-th/9807099}{{\ttfamily hep-th/9807099}}].

\bibitem{Streater:1989vi}
R.~F. Streater and A.~S. Wightman, \emph{{PCT, spin and statistics, and all
  that}}. Redwood City, USA: Addison-Wesley, 1989.

\bibitem{Haag:1992hx}
R.~Haag, \emph{{Local quantum physics: Fields, particles, algebras}}. Berlin,
  Germany: Springer, 1992.

\bibitem{Bogolyubov:1990kw}
N.~N. Bogolyubov, A.~A. Logunov, A.~I. Oksak and I.~T. Todorov, \emph{{General
  Principles of Quantum Field Theory}}. Dordrecht, Netherlands: Kluwer, 1990.

\bibitem{Kallen:1952zz}
G.~{K\"{a}ll\'{e}n}, \emph{{On the definition of the Renormalization Constants
  in Quantum Electrodynamics}}, {\emph{Helv. Phys. Acta} {\bfseries 25} (1952)
  417}.

\bibitem{Lehmann:1954xi}
H.~Lehmann, \emph{{On the Properties of propagation functions and
  renormalization contants of quantized fields}},
  \href{https://doi.org/10.1007/BF02783624}{\emph{Nuovo Cim.} {\bfseries 11}
  (1954) 342}.

\bibitem{Lowdon:2022keu}
P.~Lowdon, \emph{{Euclidean thermal correlation functions in local QFT}},
  \href{https://doi.org/10.1103/PhysRevD.106.045028}{\emph{Phys. Rev. D}
  {\bfseries 106} (2022) 045028}
  [\href{https://arxiv.org/abs/2201.12180}{{\ttfamily 2201.12180}}].

\bibitem{Lowdon:2022yct}
P.~Lowdon and O.~Philipsen, \emph{{Non-perturbative insights into the spectral
  properties of QCD at finite temperature}},
  \href{https://doi.org/10.1051/epjconf/202227405013}{\emph{EPJ Web Conf.}
  {\bfseries 274} (2022) 05013}
  [\href{https://arxiv.org/abs/2211.12073}{{\ttfamily 2211.12073}}].

\bibitem{Lowdon:2021ehf}
P.~Lowdon, R.-A. Tripolt, J.~M. Pawlowski and D.~H. Rischke, \emph{{Spectral
  representation of the shear viscosity for local scalar QFTs at finite
  temperature}}, \href{https://doi.org/10.1103/PhysRevD.104.065010}{\emph{Phys.
  Rev. D} {\bfseries 104} (2021) 065010}
  [\href{https://arxiv.org/abs/2104.13413}{{\ttfamily 2104.13413}}].

\bibitem{Lowdon:2022ird}
P.~Lowdon and R.-A. Tripolt, \emph{{Real-time observables from Euclidean
  thermal correlation functions}},
  \href{https://doi.org/10.1103/PhysRevD.106.056006}{\emph{Phys. Rev. D}
  {\bfseries 106} (2022) 056006}
  [\href{https://arxiv.org/abs/2202.09142}{{\ttfamily 2202.09142}}].

\bibitem{HotQCD:2018pds}
{\scshape HotQCD} collaboration, \emph{{Chiral crossover in QCD at zero and
  non-zero chemical potentials}},
  \href{https://doi.org/10.1016/j.physletb.2019.05.013}{\emph{Phys. Lett. B}
  {\bfseries 795} (2019) 15}
  [\href{https://arxiv.org/abs/1812.08235}{{\ttfamily 1812.08235}}].

\bibitem{Brower:2012vk}
R.~C. Brower, H.~Neff and K.~Orginos, \emph{{The M\"obius domain wall fermion
  algorithm}}, \href{https://doi.org/10.1016/j.cpc.2017.01.024}{\emph{Comput.
  Phys. Commun.} {\bfseries 220} (2017) 1}
  [\href{https://arxiv.org/abs/1206.5214}{{\ttfamily 1206.5214}}].

\bibitem{Workman:2022ynf}
{\scshape Particle Data Group} collaboration, \emph{{Review of Particle
  Physics}}, {\emph{PTEP} {\bfseries 2022} (2022) 083C01}.

\bibitem{LHCb:2017swu}
{\scshape LHCb} collaboration, \emph{{Studies of the resonance structure in
  $D^{0} \rightarrow K^\mp \pi ^\pm \pi ^\pm \pi ^\mp $ decays}},
  \href{https://doi.org/10.1140/epjc/s10052-018-5758-4}{\emph{Eur. Phys. J. C}
  {\bfseries 78} (2018) 443}
  [\href{https://arxiv.org/abs/1712.08609}{{\ttfamily 1712.08609}}].

\bibitem{Henning:1995ft}
P.~A. Henning, E.~Polyachenko and T.~Schilling, \emph{{Approximate spectral
  functions in thermal field theory}},
  \href{https://doi.org/10.1103/PhysRevD.54.5239}{\emph{Phys. Rev. D}
  {\bfseries 54} (1996) 5239}
  [\href{https://arxiv.org/abs/hep-ph/9510322}{{\ttfamily hep-ph/9510322}}].

\bibitem{Rohrhofer:2019qwq}
C.~Rohrhofer, Y.~Aoki, G.~Cossu, H.~Fukaya, C.~Gattringer, L.~Y. Glozman,
  S.~Hashimoto, C.~B. Lang and S.~Prelovsek, \emph{{Symmetries of spatial meson
  correlators in high temperature QCD}},
  \href{https://doi.org/10.1103/PhysRevD.100.014502}{\emph{Phys. Rev. D}
  {\bfseries 100} (2019) 014502}
  [\href{https://arxiv.org/abs/1902.03191}{{\ttfamily 1902.03191}}].

\bibitem{Rohrhofer:2019qal}
C.~Rohrhofer, Y.~Aoki, L.~Y. Glozman and S.~Hashimoto, \emph{{Chiral-spin
  symmetry of the meson spectral function above $T_c$}},
  \href{https://doi.org/10.1016/j.physletb.2020.135245}{\emph{Phys. Lett. B}
  {\bfseries 802} (2020) 135245}
  [\href{https://arxiv.org/abs/1909.00927}{{\ttfamily 1909.00927}}].

\bibitem{Glozman:2022lda}
L.~Y. Glozman, O.~Philipsen and R.~D. Pisarski, \emph{{Chiral spin symmetry and
  the QCD phase diagram}},
  \href{https://doi.org/10.1140/epja/s10050-022-00895-4}{\emph{Eur. Phys. J. A}
  {\bfseries 58} (2022) 247}
  [\href{https://arxiv.org/abs/2204.05083}{{\ttfamily 2204.05083}}].

\bibitem{JUWELS}
D.~Alvarez, \emph{{JUWELS Cluster and Booster: Exascale Pathfinder with Modular
  Supercomputing Architecture at Juelich Supercomputing Centre}},
  \href{https://doi.org/10.17815/jlsrf-7-183}{\emph{JLSRF} {\bfseries 7} (2021)
  A138}.

\bibitem{Boyle:2015tjk}
P.~Boyle, G.~Cossu, A.~Yamaguchi and A.~Portelli, \emph{{Grid: A next
  generation data parallel C++ QCD library}}, {\emph{PoS} {\bfseries
  LATTICE2015} (2016) 023} [\href{https://arxiv.org/abs/1512.03487}{{\ttfamily
  1512.03487}}].

\bibitem{Yamaguchi:2022feu}
A.~Yamaguchi, P.~Boyle, G.~Cossu, G.~Filaci, C.~Lehner and A.~Portelli,
  \emph{{Grid: OneCode and FourAPIs}},
  \href{https://doi.org/10.22323/1.396.0035}{\emph{PoS} {\bfseries LATTICE2021}
  (2022) 035} [\href{https://arxiv.org/abs/2203.06777}{{\ttfamily
  2203.06777}}].

\bibitem{GPTCode}
{C. Lehner \textit{et al.}}, ``{Grid Python Toolkit (GPT)}.''
  {\texttt{\href{https://github.com/lehner/gpt}{https://github.com/lehner/gpt}}}.

\bibitem{Data_DOI}
D.~Bala, O.~Kaczmarek, P.~Lowdon, O.~Philipsen and T.~Ueding, ``{Data
  publication for: \textit{Pseudo-scalar meson spectral properties in the
  chiral crossover region of QCD}}.''
  {\texttt{\href{https://doi.org/10.4119/unibi/2989968}{10.4119/unibi/2989968}}},
  (2024).

\bibitem{Nielsen:1981hk}
H.~B. Nielsen and M.~Ninomiya, \emph{{No Go Theorem for Regularizing Chiral
  Fermions}}, \href{https://doi.org/10.1016/0370-2693(81)91026-1}{\emph{Phys.
  Lett. B} {\bfseries 105} (1981) 219}.

\bibitem{Nielsen:1980rz}
H.~B. Nielsen and M.~Ninomiya, \emph{{Absence of Neutrinos on a Lattice. 1.
  Proof by Homotopy Theory}},
  \href{https://doi.org/10.1016/0550-3213(82)90011-6}{\emph{Nucl. Phys. B}
  {\bfseries 185} (1981) 20}.

\bibitem{Nielsen:1981xu}
H.~B. Nielsen and M.~Ninomiya, \emph{{Absence of Neutrinos on a Lattice. 2.
  Intuitive Topological Proof}},
  \href{https://doi.org/10.1016/0550-3213(81)90524-1}{\emph{Nucl. Phys. B}
  {\bfseries 193} (1981) 173}.

\bibitem{Follana:2003fe}
{\scshape HPQCD} collaboration, \emph{{Further improvements to staggered
  quarks}}, \href{https://doi.org/10.1016/S0920-5632(03)02610-0}{\emph{Nucl.
  Phys. B Proc. Suppl.} {\bfseries 129} (2004) 447}
  [\href{https://arxiv.org/abs/hep-lat/0311004}{{\ttfamily hep-lat/0311004}}].

\bibitem{Follana:2006rc}
{\scshape HPQCD, UKQCD} collaboration, \emph{{Highly improved staggered quarks
  on the lattice, with applications to charm physics}},
  \href{https://doi.org/10.1103/PhysRevD.75.054502}{\emph{Phys. Rev. D}
  {\bfseries 75} (2007) 054502}
  [\href{https://arxiv.org/abs/hep-lat/0610092}{{\ttfamily hep-lat/0610092}}].

\bibitem{Kaplan:1992bt}
D.~B. Kaplan, \emph{{A Method for simulating chiral fermions on the lattice}},
  \href{https://doi.org/10.1016/0370-2693(92)91112-M}{\emph{Phys. Lett. B}
  {\bfseries 288} (1992) 342}
  [\href{https://arxiv.org/abs/hep-lat/9206013}{{\ttfamily hep-lat/9206013}}].

\bibitem{Kaplan:2009yg}
D.~B. Kaplan, \emph{{Chiral Symmetry and Lattice Fermions}},  in \emph{{Les
  Houches Summer School: Session 93: Modern perspectives in lattice QCD:
  Quantum field theory and high performance computing}}, pp.~223--272, 12,
  2009, \href{https://arxiv.org/abs/0912.2560}{{\ttfamily 0912.2560}}.

\bibitem{Bar:2005tu}
O.~Bar, C.~Bernard, G.~Rupak and N.~Shoresh, \emph{{Chiral perturbation theory
  for staggered sea quarks and Ginsparg-Wilson valence quarks}},
  \href{https://doi.org/10.1103/PhysRevD.72.054502}{\emph{Phys. Rev. D}
  {\bfseries 72} (2005) 054502}
  [\href{https://arxiv.org/abs/hep-lat/0503009}{{\ttfamily hep-lat/0503009}}].

\end{thebibliography}\endgroup

\end{document}